\documentclass[]{openjournal}
\usepackage{xcolor}
\usepackage{amsmath}
\usepackage[figuresleft]{rotating}
\usepackage{booktabs}
\usepackage{graphicx}
\usepackage{etoolbox}
\usepackage{multirow}
\usepackage{array}
\usepackage{xspace}
\usepackage[colorlinks=true, linkcolor=blue, citecolor=blue, urlcolor=cyan]{hyperref}
\usepackage{orcidlink}

\defcitealias{nesvorny2023}{N23}

\newcommand{\code}[1]{\texttt{#1}}

\shorttitle{Ejection Velocities of ISOs}
\shortauthors{Albrow et al.}

\begin{document}

\title{\vspace{-0.8cm}The ejection velocities of interstellar objects signpost their progenitor system architectures\vspace{-1.5cm}}

\author{Leah Albrow\,\orcidlink{0009-0003-4251-2821}$^{1,2}$, Michele T. Bannister\,\orcidlink{0000-0003-3257-4490}$^{1}$,
John C. Forbes\,\orcidlink{0000-0002-1975-4449}$^{1}$,
David Nesvorn{\'y}\,\orcidlink{0000-0002-4547-4301}$^{3}$
}
\affiliation{
${^1}$School of Physical and Chemical Sciences--Te Kura Mat\=u, University of Canterbury, Christchurch 8140, New Zealand\\
${^2}$Department of Earth, Atmospheric, and Planetary Sciences, Massachusetts Institute of Technology, 77 Massachusetts Avenue, Cambridge, MA
02139, USA\\
$^{3}$Department of Space Studies, Southwest Research Institute, 1301 Walnut St., Suite 400, Boulder, CO 80302, USA}
\thanks{$^*$E-mail: \href{lalbrow@mit.edu}{lalbrow@mit.edu}}

\begin{abstract}
Interstellar objects (ISOs) ejected from planetary systems carry kinematic signatures of their formation environments. 
The properties of these velocity distributions govern the ISOs’ propagation and dynamical evolution in the Galactic potential.
We investigate how planetary system architecture influences ISO production during post-gas-disc dynamical instabilities using N-body simulations.
We explore the ISO production outcomes of 2461 randomly-generated systems, spanning system masses of 300-800 M$_\oplus$ and multiplicities of 3-7 planets.
Integrating planets in a disc of test particles for 10 Myr, we find that the evolving systems can be broadly divided into two distinct classes based on their initial architectures.
\textit{Catastrophic systems} are characterised by high-multiplicity and orbitally-compact architectures, or by having high-mass planets in systems with large mass asymmetries. 
These systems eject a high fraction of planetesimals (median 59\%) and, depending on the ejection pathway, eject ISOs at high speeds (median $2.9$ km s$^{-1}$). 
Alternatively, \textit{quiet systems} have lower masses and multiplicities, and do not undergo significant orbital changes, but are still able to eject a median of 28\% of planetesimals at low velocities (median $1.6$ km s$^{-1}$).
This points to different ejection pathways, involving either violent global instabilities or more gradual, diffusive processes.
We find the ejection velocities of ISOs are typically low, on the order of a few km s$^{-1}$.
Although ISOs will then experience dynamical heating as they orbit the Galaxy, their velocity distributions retain signatures of their progenitor systems’ architectures and histories, pointing to the future use of ISOs in Galactic archaeology.

\end{abstract}


\section{Introduction}\label{sec:intro}

Interstellar objects (ISOs) are formed in protoplanetary discs and are subsequently unbound from their host star through a variety of proposed dynamical processes \citep[e.g.][]{bannister2019}.
Once unbound, they traverse the Galaxy, with their long-term evolution driven by their initial ejection velocities.
As ISOs orbit within the Galactic potential, differences in ejection velocities stretch unbound populations into tidal streams. 
\citet{Forbes2024} demonstrated that stream properties—size, density structure, and kinematics—are shaped primarily by the initial velocity dispersion, with secondary contributions from Galactic perturbations.
This dispersion is set by the initial spread in ejection velocities. 
Subsequent dynamical heating occurs from non-axisymmetric perturbations in the Galactic potential.
Understanding how ISOs are ejected and their associated velocities is therefore crucial for studying their distribution throughout the Galaxy.

ISOs must first be produced before their Galactic dynamics matter.
Planetesimal formation appears ubiquitous in young protoplanetary discs, the mechanisms for which are outlined in a recent review by \citet{pp7}.
The frequency at which planetesimals and dust grains are ejected during the gas-disc phase remains unconstrained.
\citet{eriksson2021} find that, in planetesimal disc simulations which include effects such as gas drag and mass ablation to simulate the presence of the gas disc, 15–40\% of planetesimals formed at the edges of planetary gaps are ejected.
Further work is needed to understand how gas drag and disc turbulence affect planetesimal ejection rates during this phase.

In this work we focus on systems after gas disc dispersal, by processes such as magnetic winds, turbulence, stellar encounters, or external photo-evaporation \citep{hasegawa2022, alexander2014}, leaving a population of giant planets, and planetesimals.
After disc dispersal, gravitational scattering between planetesimals and the system’s massive planets becomes a dominant dynamical process. 
Gravitational scattering can drive planetary migration \citep{Levison2007}, and in the case of inward migration, scatters planetesimals onto larger orbits.
In the Solar System, gas dispersal has been linked to the onset of a large-scale dynamical instability \citep{liuEarlySolarSystem2022}.
Because most modelling efforts focus on retained populations, either to reproduce observed exoplanet architectures or to match the precisely measured small-body populations of the Solar System \citep[e.g.][]{bannister2018, pfalzner2024}, ejected material has received far less attention.
For example, in a typical Solar System simulation from  \citet{nesvorny2023}, 93.4\% of planetesimals initially placed between 4 and 30 au were ejected over the 4.5 Gyr of the simulation.
Given the broad diversity of exoplanetary systems, whether this high ejection fraction is typical remains unclear, though this has long been used as a suitable initial assumption \citep[e.g.][]{stern1990,moro-martin2009,Raymond2020,lintott2021}.

Beyond instability-driven ejections, several other mechanisms may contribute to the ISO population.
Since most ISOs are thought to be ejected by early-system processes, the birth cluster environment likely plays a significant role in their dynamical evolution.
Stellar flybys, common in dense stellar environments \citep{malmberg2011}, can perturb outer disc populations and eject ISOs with typical velocities between $0.5$ and $2$ km s$^{-1}$ relative to the host star \citep{Pfalzner2021}. 
Such encounters may also trigger planetary instabilities that further enhance ISO production.

Finally, evolved stars are hypothesised to produce ISOs.
A small wave of ISOs may be gently unbound during the asymptotic giant branch (AGB) phase of stellar evolution \citep[e.g.][]{veras2014, levine2023}.
According to ejection velocities reported in \citet{Pfalzner2021} (based on unpublished work from \citet{veras2014}), these objects have mean ejection velocities below $0.5$ km s$^{-1}$.

Dynamical instabilities are an underexplored yet significant mechanism for releasing large fractions of a system’s planetesimals as ISOs.
\citet{morbidelliDynamicalEvolutionPlanetary2018} outlined several potential triggers for global instabilities, including destabilization following disc dispersal, the formation of planets in close proximity (within a few Hill radii), capture into orbital resonances, secular chaotic evolution, tidal interactions with the host star, and external perturbations.
Such instabilities appear to be common.
For example, \citet{izidoro2017} found that, to reproduce the observed Kepler sample, more than 75\% of multi-planet systems hosting super-Earths must have undergone a dynamical instability after gas dispersal.
Additional observational evidence comes from the discovery of high-eccentricity hot Jupiters, whose in-situ formation is considered unlikely \citep{dawson2018}.
Significant dynamical evolution thus likely produced these planets, requiring either a distant perturber or planetary companion.

\begin{figure*}
    \centering
    \includegraphics{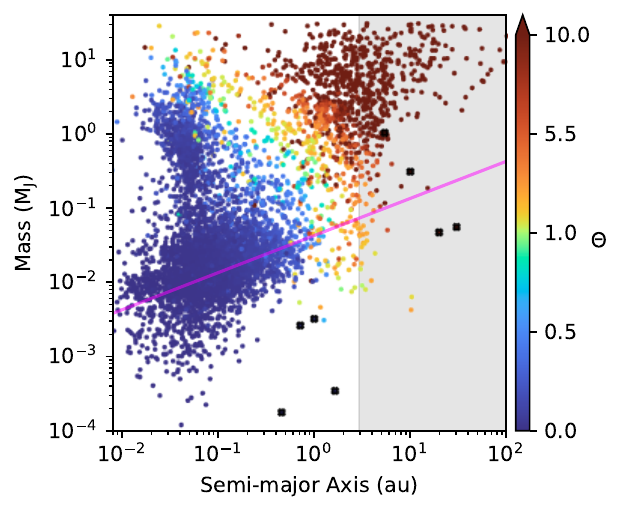}%
    \includegraphics{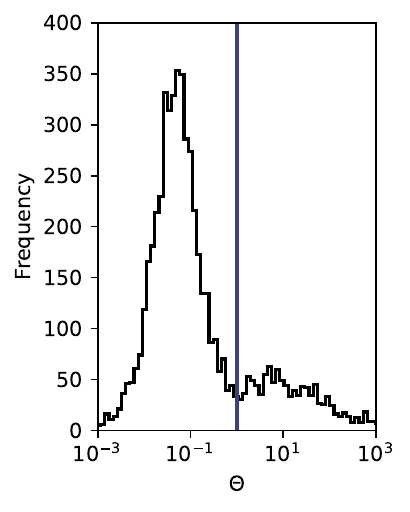}
    \caption{
    \textbf{Left}: Safronov number for the population of known exoplanets, following Fig.~1 of \citet{laughlin2017}.  
    Only planets with measured masses and radii are shown (data from the NASA Exoplanet Archive, 22 Nov 2025). 
    Solar System planets are plotted as larger symbols with black outlines.
    The pink line indicates a radial-velocity semi-amplitude precision of 1 m s$^{-1}$ \citep{lovis2010}, indicating the approximate detection threshold for RV surveys.
    The grey shaded region marks the Kepler primary mission duration, serving as a proxy for transit-survey sensitivity.  
    \textbf{Right}: distribution of $\Theta$ for the exoplanets in the left panel. 
    Most known exoplanets have $\Theta \ll 1$, implying that their dynamical interactions are collision-dominated rather than ejection-dominated.
    }
    \label{fig:safronovscatter}
\end{figure*}

The impact of dynamical instabilities on planetesimal fates is the central focus of this work.
A key quantity governing these outcomes is the Safronov number,
\begin{equation}
    \Theta = \frac{v^2_{\text{esc}}}{v^2_{\text{c}}} = \frac{2 M_p}{M_\ast}\frac{a_p}{R_p},
\end{equation}
which characterizes the ratio of ejections to planet–planetesimal collisions \citep{safronov1972}.
Here $v_{\rm esc}$ is the planet’s escape speed and $v_{\rm c}$ is its circular orbital velocity around a star of mass $M_*$.
A planet with $\Theta \gtrsim 1$ typically ejects planetesimals during close encounters rather than accreting them.
In the case where the Safronov number is less than one, collisions are more likely. 
\citet{laughlin2017} argue that the existence of ISOs requires a population of exoplanets capable of efficiently ejecting material, that is, planets with a Safronov number greater than or equal to one. 
As illustrated in \autoref{fig:safronovscatter}, the currently known exoplanet sample is strongly biased: short-period, large-radius planets are overrepresented, while long-period intermediate-mass planets remain under-detected.
A substantial population of efficient planetesimal ejectors may lie in the bottom-right regions of the left panel, where current detection methods are incomplete.
The right panel of \autoref{fig:safronovscatter} shows the majority of known exoplanets cannot efficiently eject ISOs.
Because our interest lies in ISO ejection, we focus on systems dominated by planets with Safronov numbers greatly exceeding unity.
Rather than attempting to correct for exoplanet observational biases, we instead investigate ensembles of randomly generated planetary system architectures \citep[see e.g.,][for a similar approach]{Bhaskar:2025}, with an emphasis on the outer planets.

We begin by describing our N-body simulations (\autoref{sec:integrations}), then examine the dynamical evolution of a representative system to illustrate key ejection mechanisms (\autoref{sec:individual-system}).
Statistical analysis of our full simulation ensemble (\autoref{sec:ensemble}) reveals different pathways to ejection based on planetary system architectures, and trends in the inclinations of ejected ISOs, which are directly relevant for modelling ISO tidal streams \citep{Forbes2024}. 
We then use clustering methods (\autoref{sec:clusters}) to find trends in the system architectures of systems with similar ejection velocity distributions.
Finally, in \autoref{sec:discussion} we consider the broader implications of our findings and outline directions for future work. 

\section{Planetary System Integrations}
\label{sec:integrations}

\subsection{Dynamical Instability Simulations}

We simulated the evolution of planetary systems embedded within discs of massless particles, exploring a wide range of system architectures. 
These simulations were conducted using the \code{rebound} Python package \citep{rebound}. 
We used the \code{MERCURIUS} hybrid integrator, which employs a symplectic Wisdom-Holman integrator (\code{WHFast}) for well-separated particles and switches to a high-order integrator (\code{IAS15}) during close encounters \citep{wisdom1991, reinwh2015, rein2019}.
The threshold for switching to close-encounter mode was set at three Hill radii from the planets.

We focused on ISO production during dynamical instabilities rather than long-term stability, so we integrated systems for 10 Myr.
Given the high masses of the simulated systems, which drive stronger gravitational perturbations and shorter dynamical timescales, this duration was sufficient for numerous instabilities to develop.
Simulation states were stored in the \code{Simulationarchive} format \citep{simulationarchive}, allowing efficient storage and processing of particle trajectories.

A total of 2,500 planetary systems were generated as follows.
The total mass of planets in the systems took values of 300, 400, 500, 600, 700, and 800~M$_\oplus$ (for comparison, the Solar System’s planetary mass is $\sim$335~$M_\oplus$).
The initial multiplicity (number of planets) ranged from 3 to 7.
100 systems were created for each combination of multiplicity and total system mass.
Several simulations did not finish running due to computational limitations, leaving a total of 2461 completed simulations.
Individual planet masses were drawn from a symmetric Dirichlet distribution, which produces $n$ positive values (where $n$
is the system multiplicity) summing to unity.
We used a concentration parameter of $\alpha=1$, corresponding to a uniform distribution over all possible mass partitions.
We then scaled these fractions by the system's total planetary mass to obtain individual planet masses.
The star was fixed at one solar mass.

\begin{figure*}
    \centering
    \includegraphics{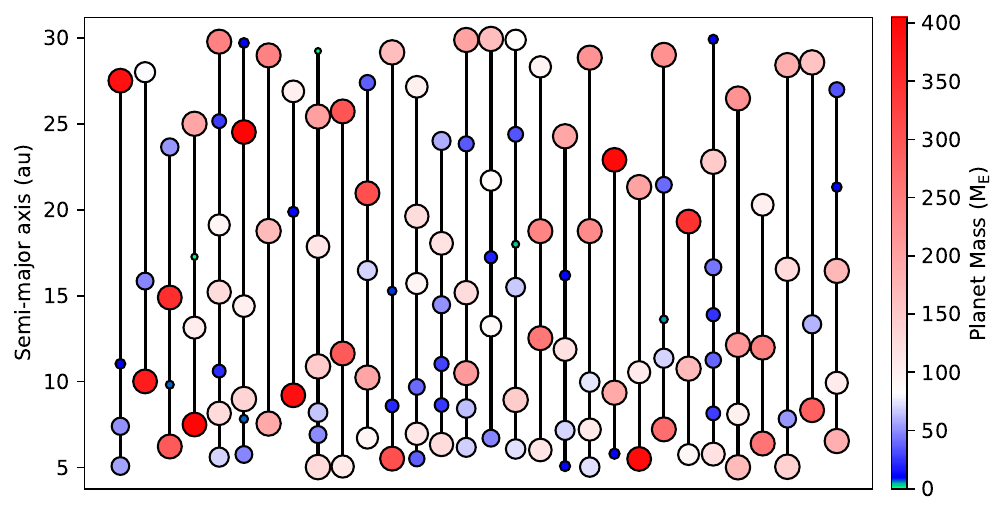}
    \caption{Thirty of the generated initial system configurations.
    Colour encodes planetary mass, indicating approximate compositional categories.
    Red planets are larger than 80 M$_\oplus$ (gas giants), and blue planets are from 5-80 M$_\oplus$ (gas and ice giants). 
    The green planets are below 5 M$_\oplus$, indicating terrestrial planets. 
    Symbol sizes are proportional to planetary radii calculated from \autoref{eq:massradius}.}
    \label{fig:examplesystems}
\end{figure*}

For each system, planetary semi-major axes were randomly drawn, with equal probability, from 10,000 candidate positions uniformly distributed between 5 and 30 au. 
Planets were assigned to positions randomly without regard to mass.
We then masked regions within eight Hill radii of each placed planet to prevent overlaps.
In cases where no valid placements remained -- particularly in higher-mass systems -- the exclusion radius was gradually reduced until all planets could be placed.
For the highest mass systems, the minimum number of Hill radii was decreased to 5.2. 
Examples of the generated planetary systems are shown in \autoref{fig:examplesystems}.
Planetesimals were placed in a similar fashion, but with a reduced minimum of four Hill radii separation from the planets.
This sampling scheme yields, on average, both a planet and a planetesimal surface density profile $\propto r^{-1}$, though with considerable system-to-system scatter for the planets.

No additional stability constraints were imposed during system generation.
Instead, systems were allowed to evolve naturally and were expected to undergo dynamical instability over the 10 Myr integration due to their high mass and compactness.
Planets were treated as point masses during the integrations, but were later assigned estimated radii using the mass–radius relationship from \citet{massradius}:
\begin{equation}\label{eq:massradius}
    R = \begin{cases}
        1.02~M^{0.27} & \text{if } M < 4.37, \\
        0.56~M^{0.67} & \text{if } 4.37 \leq M < 127, \\
        18.6~M^{-0.06} & \text{if } M \geq 127,
    \end{cases}
\end{equation}
where $M$ and $R$ are in Earth units.
This estimated radius is needed to calculate some of the system architecture parameters later introduced in \autoref{tab:system-params}.

Each system had 10,000 massless planetesimals as a proxy for a mildly heated debris disc, extending from 5 to 40 au from the central star as described above. 
Orbital parameters for both planets and planetesimals were initialized from the following distributions: eccentricities drawn uniformly with $0 \leq e \leq 0.1$, inclinations between $0 \leq i \leq 0.1$ radians, and true anomaly ($f$), argument of periapsis ($\omega$), and longitude of ascending node ($\Omega$) drawn uniformly between $0$ and $2\pi$. 
These ranges were chosen to represent a warm, but not yet dynamically excited, disc.
Because the planetesimals were massless and thus could not self-excite, slightly more dynamically excited initial conditions were adopted to approximate natural disc heating.
All planetesimals were treated as point particles in the integrations. 

Planets were removed from the simulation if they satisfied either of the following conditions: (1) $e > 1$ and $r > 200$ au or (2) $a>1000$ au and $e>0.95$. 
The radial distance threshold prevents premature removal of planets still interacting with the planetesimal disc.
To ensure complete tracking of all ejection events, test particles were never removed.
At the end of the simulation, ISOs were identified as all test particles with $e > 1$.
We calculated their barycentric ejection velocities ($v_\infty$) for analysis.

Because planets were treated as point masses\footnote{We examined the final eccentricity distributions of the simulated planets, which were slightly broader than those observed — consistent with expectations from the point-mass approximation, which enhances the dynamical heating of giant planets.}, they may yield unrealistically large ejection velocities.
In reality, planetesimals approaching too closely would either collide with the planetary surface or undergo tidal disruption, depending on their internal strength.
To account for physical collisions, we applied an upper velocity cutoff as follows: the maximum possible heliocentric $v_\infty$ following an encounter with a planet is $2\sqrt{1+\sqrt{2}} \approx 3.1$ times the planet's circular velocity, assuming the planet is on a nearly circular orbit. 
This limit corresponds to the case where a planetesimal can approach arbitrarily close to a point-mass planet ($\Theta \to \infty$).
In the opposite limit where $\Theta \to 0$, the maximum possible $v_\infty$ approaches zero. 
For Safronov numbers of order 10--100, we find numerically that the maximum $v_\infty/v_c \approx 2$. 
We therefore filter out $v_\infty$ values above
\begin{equation}
    v_{\rm max} = 2 \max(v_p),
\end{equation}
where $\max(v_p)$ is the circular velocity of the innermost planet in the system.
Since our initial minimum semi-major axis is 5 au, this imposes a maximum velocity of $\sim 27$ km s$^{-1}$.
We checked our systems for inward migration of the planets, and adjusted this limit if any planets had final $a$ < 5 au.
This filter excludes particles that would have collided with the planet's physical surface in a more realistic simulation.

\subsection{Solar System Simulation}

We compared our results to a high-resolution Solar System evolution simulation by \citet{nesvorny2023} (hereafter \citetalias{nesvorny2023}). 
This simulation included additional physical processes: planetary migration, instability, stellar encounters, and Galactic tides.
The Solar System's birth cluster was not a part of this simulation.
The \citetalias{nesvorny2023} simulation spans 4.5 Gyr and defines ejected particles as those reaching a barycentric distance of 500,000 au.
The initial population consisted of $10^6$ particles distributed between 4 and 30 au, with a surface density scaling $\propto r^{-1}$.
Bound particles which remained in the Oort Cloud were excluded from our analysis.

The positions and velocities of the particles reaching 500,000 au were recorded, along with the epoch at which they crossed this radius.
To allow comparison with our \code{rebound} simulations, where particles are flagged as ``ejected’’ at the time they are unbound, we calculated the time of ejection for the \citetalias{nesvorny2023} particles.
For a hyperbolic orbit, the mean anomaly $M$ is given by
\begin{equation}
    M = \sqrt{\frac{\mu}{-a^3}} (t - \tau),
\end{equation}
where $\mu$ is the gravitational parameter ($GM_*$), $t$ is the time at which the particle reaches 500,000 au, and $\tau$ is the time of the last perihelion passage. 
The orbital elements were computed from each particle’s position and velocity.
From these, we extracted $\tau$ and the hyperbolic excess velocity, $v_\infty = \sqrt{-\mu/a}$.

\section{Results}

\subsection{Individual System Analysis}
\label{sec:individual-system}

\begin{figure}
    \centering
    \includegraphics {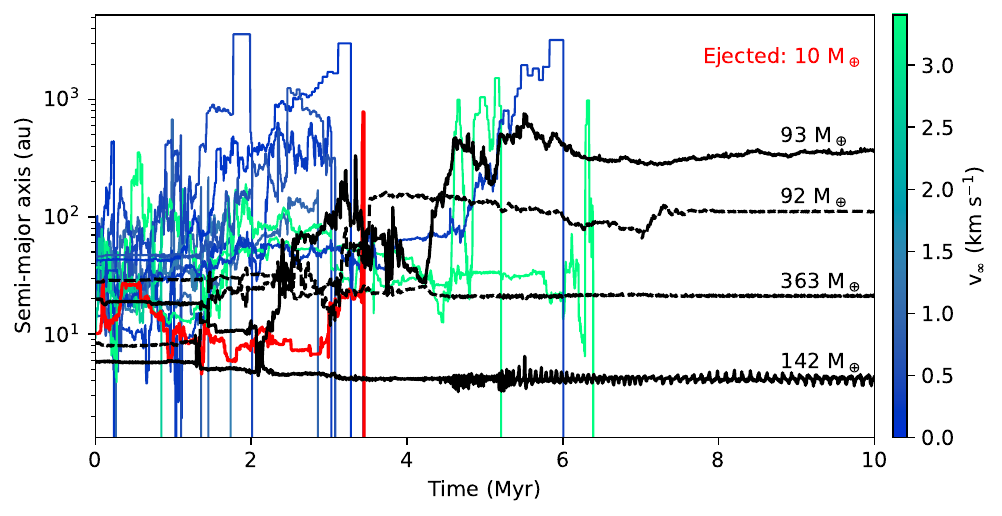}
    \caption{Semi-major axis time series for planets and 16 randomly chosen ejected planetesimals in the example system.
    Colours from blue to green indicate ejection velocity magnitude for ISOs.
    The red curve shows the ejected 10 M$_\oplus$ planet; black curves show planets that remain bound at 10 Myr.} 
    \label{fig:time-series-individual}
\end{figure}

Individual system evolution reveals physical mechanisms that population statistics alone cannot capture.
We first present detailed results from a representative example simulation, illustrating the mechanisms we then quantify statistically in \autoref{sec:ensemble}.
The example simulation contains five planets with a total initial planetary mass of 700 M$_\oplus$. 
The system has a median $v_\infty=2.31$ km s$^{-1}$, and the fraction of ejected planetesimals is 0.664.

\begin{figure}
    \centering
    \includegraphics{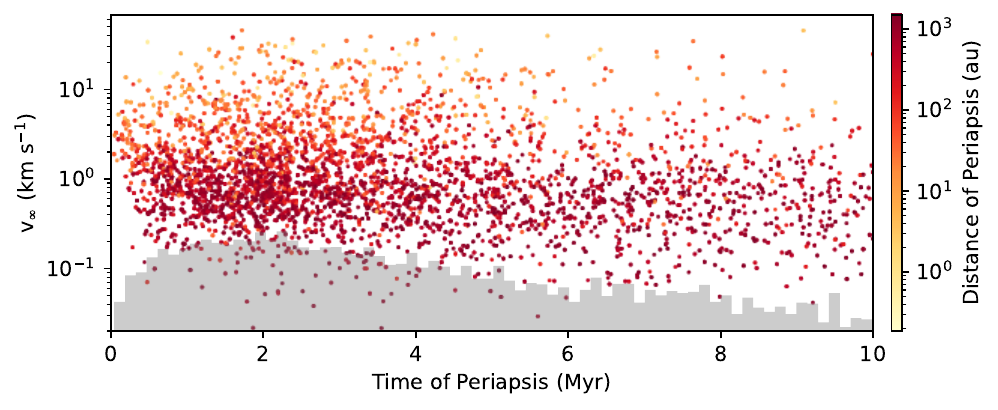}
    \caption{Ejection time, perihelion distance, and velocity for all ISOs in the example system. 
    Early ejections ($t < 2$ Myr) from inner regions show the highest velocities.
    Ejection velocity decreases with perihelion distance.
    The grey histogram shows the ejection flux, which peaks at 2 Myr after simulation initialisation, and then steadily decreases.
    }
    \label{fig:semi-ax-vinf}
\end{figure}

\autoref{fig:time-series-individual} shows the time evolution of the planetary semi-major axes alongside 16 randomly-selected ejected particles.
The system undergoes a catastrophic instability early in its evolution. 
Most planets experience significant orbital changes, and one 10 M$_\oplus$ planet is ejected.
The planet trajectories also indicate that mass asymmetry drives instability outcomes, with the largest planets remaining stable, and the smaller planets ejected or scattered onto high semi-major axis orbits.
ISOs are ejected through two main pathways. 
Some experience single strong scattering events that directly impart hyperbolic velocities, while others undergo gradual outward diffusion in semi-major axis before gentle unbinding.
The majority of the shown ISOs are ejected in the first two Myr of the simulation, as seen in the ejection flux shown in the gray histogram of \autoref{fig:semi-ax-vinf}.
The early flux increase may reflect either the planetary instability or relaxation of initially unstable planetesimal orbits.
This plot also shows which regions of the planetesimal disc contribute most to the ejected population, with the majority of ejected objects having a perihelion exceeding that of the planets at any given time.
Planetesimals located farther from the star at the time of ejection are ejected with lower $v_\infty$, as expected for weakly bound, slower-moving orbits.
Low-velocity, distant ejections become more dominant later in the integration.
This pattern suggests that early ejections are dominated by close encounters near the planets, while later ones arise from more gradual diffusion of distant planetesimals.

\begin{figure}
    \centering
    \includegraphics{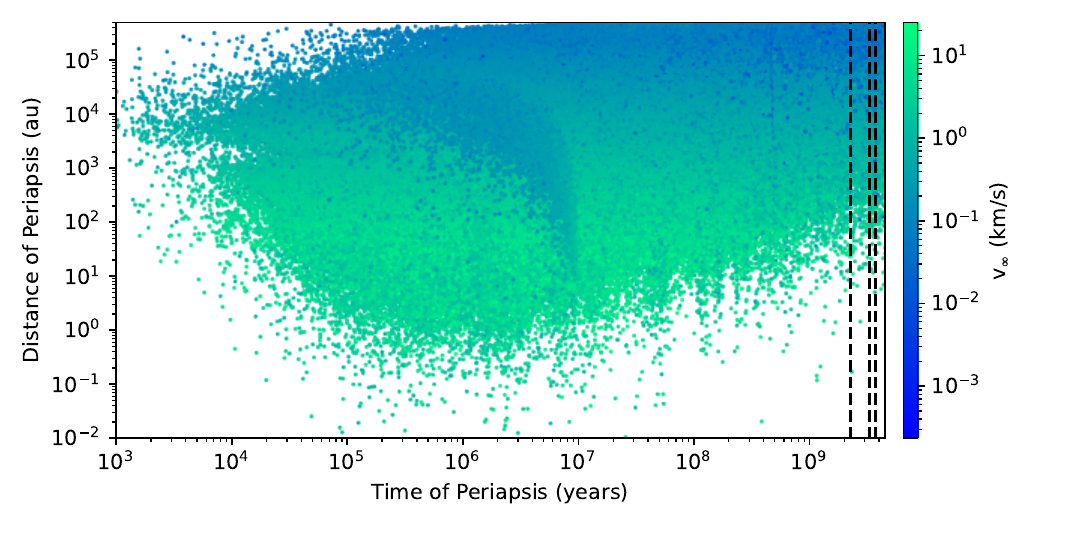}
    \caption{
    Temporal evolution of ISO ejection velocities in the Solar System simulation of \citetalias{nesvorny2023}.
    Each point shows an ejected particle’s $v_\infty$ at the time of ejection, calculated from its orbital elements at 500,000 au and traced back to perihelion passage time.
    Colour indicates perihelion distance, while vertical dashed lines correspond to strong stellar flyby events that perturb the system. 
    Additional vertical features may indicate further stellar encounters.
    }
    \label{fig:perihelion-scatter}
\end{figure}

The system's velocity distribution evolves as inner planetesimals are preferentially depleted early. 
The median $v_\infty$ peaks at 3.0 km s$^{-1}$ during the initial instability ($\sim 0.25$ Myr after the start of the simulation), then declines to a plateau of 0.9 km s$^{-1}$ by 10 Myr.
This behaviour parallels the Solar System evolution seen in \citetalias{nesvorny2023}, though with lower velocities (the Solar System simulation has $8.0$ km s$^{-1}$ peak and $1.0$ km s$^{-1}$ plateau).
A diffusion timescale is required for particles to migrate onto orbits large enough to be gently ejected at low velocities.
This is evident in \autoref{fig:perihelion-scatter}, where the maximum perihelion distance increases throughout the 4.5 Gyr integration.
Particles were removed at 500,000 au barycentric distance; the upper envelope in perihelion distance reflects this removal threshold combined with orbital eccentricity.
Additionally, these ejections are the slowest ($< 10^{-3}$ km s$^{-1}$), leading to a gradual build-up of the low-velocity tail of the distribution.
The low-velocity tail is also sensitive to physical effects not included in our simulations.
We omit both stellar flybys and Galactic tides: the former can generate slow-moving ISOs \citep{Pfalzner2021}, and the latter plays an important role in shaping Oort-cloud dynamics \citep{fernandezFormationOortCloud1997}, where planetesimals’ Keplerian velocities are much lower than in the planetary region.
Finally, our 10 Myr integrations may be too short for planetesimals originating near the outer disk edge (40 au) to diffuse onto the high-eccentricity, distant orbits required for such low-velocity ejections.

\begin{figure}
    \centering
    \includegraphics{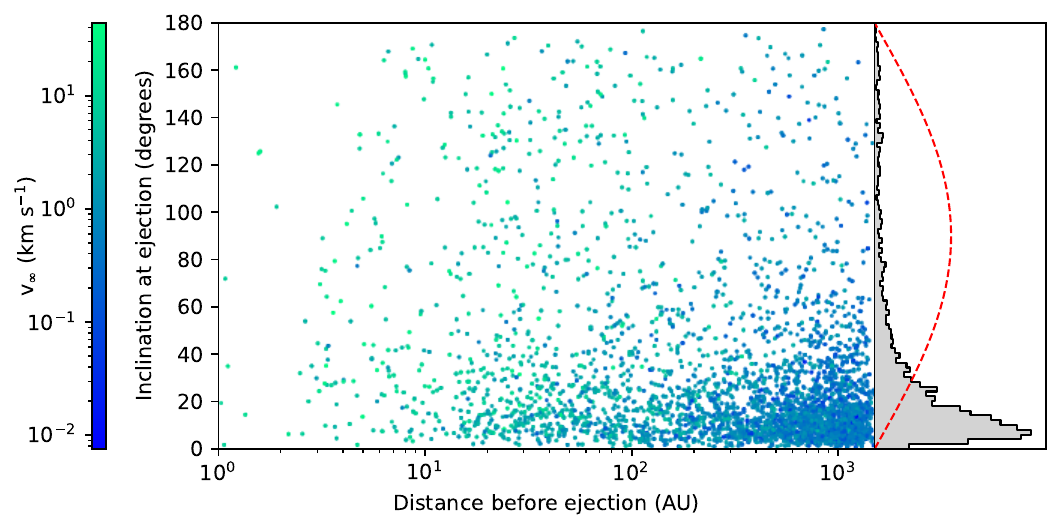}
    \caption{Inclinations of ISOs ejected from the example system. 
    The fraction of low inclination ejections increases with respect to distance from the star.
    Most ejections remain at low inclination.
    The red dashed line shows the distribution expected for isotropic ejections.}
    \label{fig:inc-scatter}
\end{figure}

\begin{figure}
    \centering
    \includegraphics{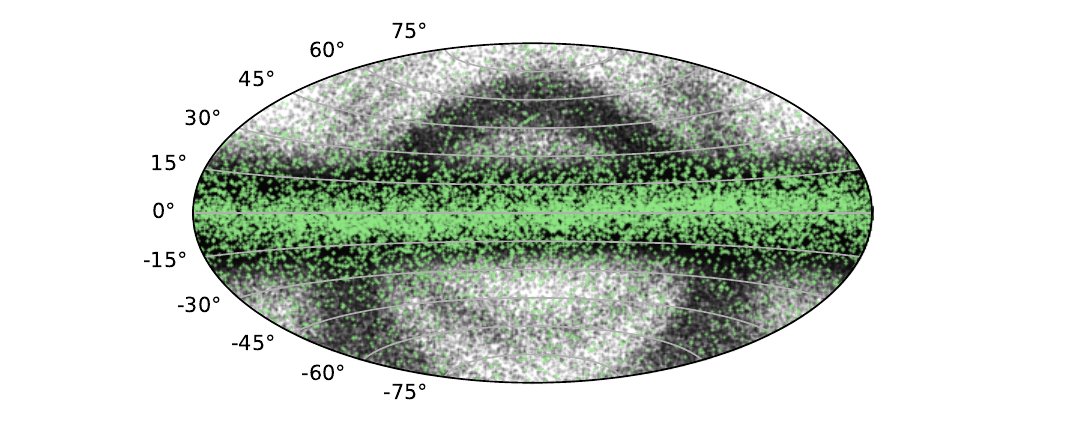}
    \caption{Galactic latitude and longitude of ejected particles. Black points are from \citetalias{nesvorny2023} over the whole 4.5 Gyr of the simulation, and green points show our example simulation after 10 Myr. 
    The majority of ejected ISOs are in the ecliptic, with the particles from \citetalias{nesvorny2023} showing the effects of integration in the Galactic tide and flybys. Particles from both simulations decrease in density at greater latitudes.}
    \label{fig:latitude}
\end{figure}

The inclination distribution of ejected ISOs, shown in \autoref{fig:inc-scatter}, provides a kinematic fingerprint of the underlying ejection mechanism. 
The distribution is strongly anisotropic and prograde-dominated, with most ISOs ejected at low inclinations.
The absence of planetesimal self-stirring in our simulations likely accentuates this planar bias. 
Self-gravitating debris discs develop enhanced inclination dispersion through collective effects. 
This would seed a broader range of initial inclinations, producing more isotropic high-velocity ejections.
Direct comparison of the ejection inclination destructions from our system with \citetalias{nesvorny2023} is difficult.
The Galactic tide significantly alters the non-radial component of the velocity of ejected objects from the ejection point to the measurement at 500,000 au. 
To compare the inclination distributions, we instead plot the Galactic latitude and longitude of the ISO ejections from our system and \citetalias{nesvorny2023}, centred on the barycentre of the system, shown in \autoref{fig:latitude} and oriented around the invariant plane of the system.
Both systems show strong concentration toward the invariant plane, with the \citetalias{nesvorny2023} particles reproducing the expected pattern from the effects of the Galactic tide. 

The individual system analysis demonstrates that ISO production efficiency and ejection kinematics vary systematically with planetary architecture and evolutionary phase. 
For ensemble-level analysis, we therefore focus on two complementary metrics: the ejected fraction (total ISOs produced relative to the initial planetesimal population) and the median ejection velocity $v_\infty$. 
The ejected fraction quantifies system-to-system variability in ISO production efficiency, while median $v_\infty$ characterizes the typical kinematic signature. 
We use the median rather than mean because it is robust to rare high-velocity outliers from close encounters.

\subsection{Simulation Ensemble Analysis}
\label{sec:ensemble}

\begin{figure}[ht]
    \centering
    \includegraphics{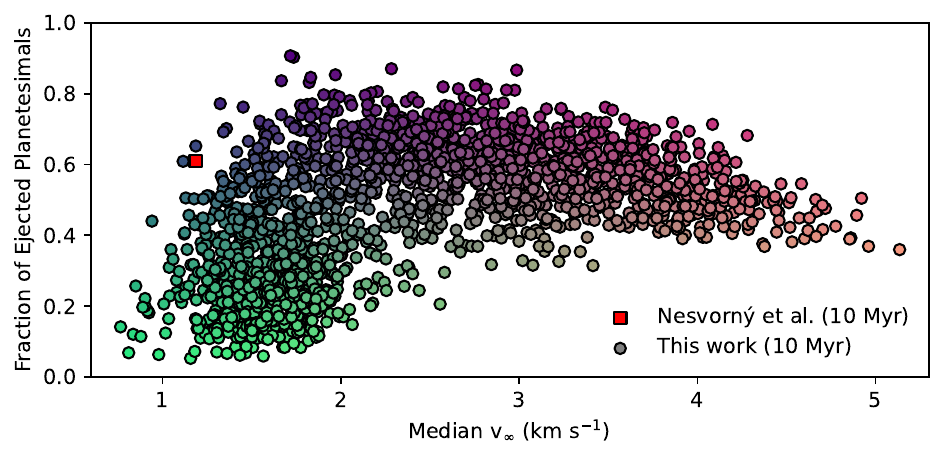}
    \caption{Summary statistics for all 2,461 simulated systems: fraction of planetesimals ejected versus the median $v_\infty$ of ejected particles.
    Colour encodes position in this parameter space for identification in subsequent figures.
    The red square marks the Solar System value from \citetalias{nesvorny2023} including all particles that have reached 500,000 au, and that have pericentre passage times earlier than 10 Myr. Over time, the \citetalias{nesvorny2023} point would move up and to the left as slower particles unbound from the Oort cloud contribute to the velocity distribution.}
    \label{fig:all-sims-scatter}
\end{figure}

The ensemble of 2,461 completed simulations reveals bimodality in ISO production outcomes.
In the median $v_\infty$–ejection fraction plane (\autoref{fig:all-sims-scatter}), systems segregate into an L-shaped distribution comprising two distinct branches.
The vertical lobe spans a wide range of ejection fractions ($\lesssim 0.7$) while maintaining relatively modest median velocities ($\sim 1$–$2$ km s$^{-1}$). 
The horizontal lobe, by contrast, clusters at elevated ejection fractions ($\gtrsim 0.4$) and exhibits higher median velocities ($\sim 2$–$5$ km s$^{-1}$).
This bimodality reflects two fundamentally different modes of planetesimal disc dispersal: gentle, protracted ejection driven by weak secular perturbations, and violent ejection accompanying global dynamical instabilities. 

\begin{figure}
    \centering
    \includegraphics{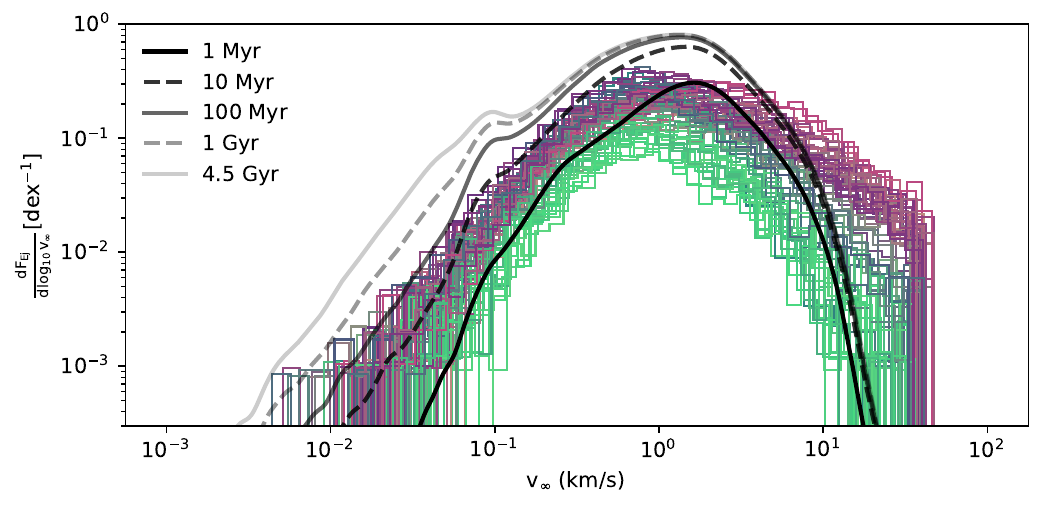}
    \caption{The $v_\infty$ distributions for 80 randomly selected simulations, coloured according to their positions in \autoref{fig:all-sims-scatter}.
    Black and gray curves show the Solar System ejection velocity distributions from \citetalias{nesvorny2023} at different times. For example, ``10 Myr'' refers to ejected particles from \citetalias{nesvorny2023} that reach 500,000 au with pericentre passage times $\tau < 10$ Myr.}
    \label{fig:all-sims-hist}
\end{figure}

The full velocity distributions for 80 randomly selected simulations are shown in \autoref{fig:all-sims-hist}, coloured consistently with their positions in the median $v_{\infty}$-ejection fraction plane.
For comparison, we also include the ejection-velocity distributions from the \citetalias{nesvorny2023} Solar System integration at several evolutionary snapshots.
At 10 Myr (our maximum integration time), the Solar System’s $v_\infty$ distribution has a similar peak and width to those of our systems, most closely resembling the distributions in the lower-left region of \autoref{fig:all-sims-scatter} (green histograms).
The higher-velocity tail in our simulations likely reflects two factors: the omission of collision handling (which would otherwise damp extreme scattering outcomes) and the higher mass budgets in our models compared to the Solar System, both of which increase the frequency and energy of strong scattering events.
By 100 Myr, the \citetalias{nesvorny2023} Solar System simulation develops a pronounced low-velocity tail.
This less-numerous late-emerging population arises from the gradual stripping of the extended Oort Cloud through a combination of weak planetary perturbations, Galactic tides, and passing stars.
Our simulations, which omit these external perturbations and conclude at 10 Myr, do not reproduce this feature.
Its absence highlights a key difference in scope: our ISO populations represent the bodies produced by dynamical instability within planetary systems, not those generated by longer-term, stripping processes.

\begin{figure}
    \centering
    \includegraphics{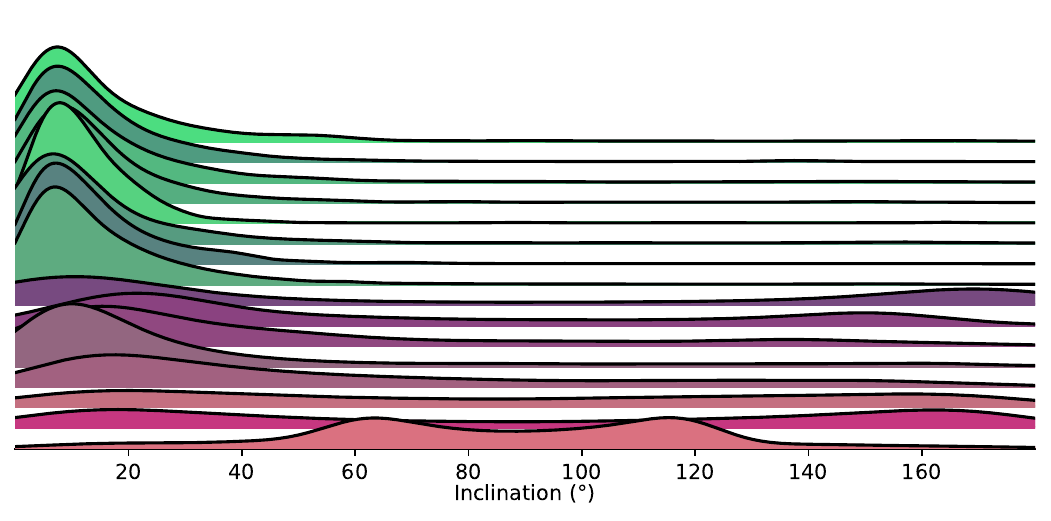}
    \caption{Inclination distribution of ejected ISOs of 25 randomly selected simulations, coloured according to their positions in \autoref{fig:all-sims-scatter}. 
    Green-coloured systems predominantly have prograde low-inclination ejections.
    The systems with the highest ejection velocities display more extreme inclination distributions. 
    }
    \label{fig:all-systems-inc}
\end{figure}

We also examined the inclination distribution across the simulation ensemble.
\autoref{fig:all-systems-inc} shows 25 systems selected from across the parameter space.
There are major structural differences in the inclination distributions depending on their position in the median $v_\infty$-ejection fraction plane.
The lowest ejection fraction systems (green) primarily have prograde ejections, with relatively low inclinations, displaying strongly asymmetric distributions
The intermediate (purple) systems have broader distributions. 
The majority of ejections remain prograde, with higher typical inclinations than green systems. 
Many purple systems show bimodal distributions, with a second peak of high inclination (prograde) ejections.
Orange systems often show bimodal distributions with peaks near 90$^\circ$. 
This reflects increasing energy of scattering events as systems move up and across the median $v_\infty$-ejection fraction plane.
None of the systems produce isotropic ejection distributions.

\begin{sidewaysfigure}
    \centering
    \includegraphics{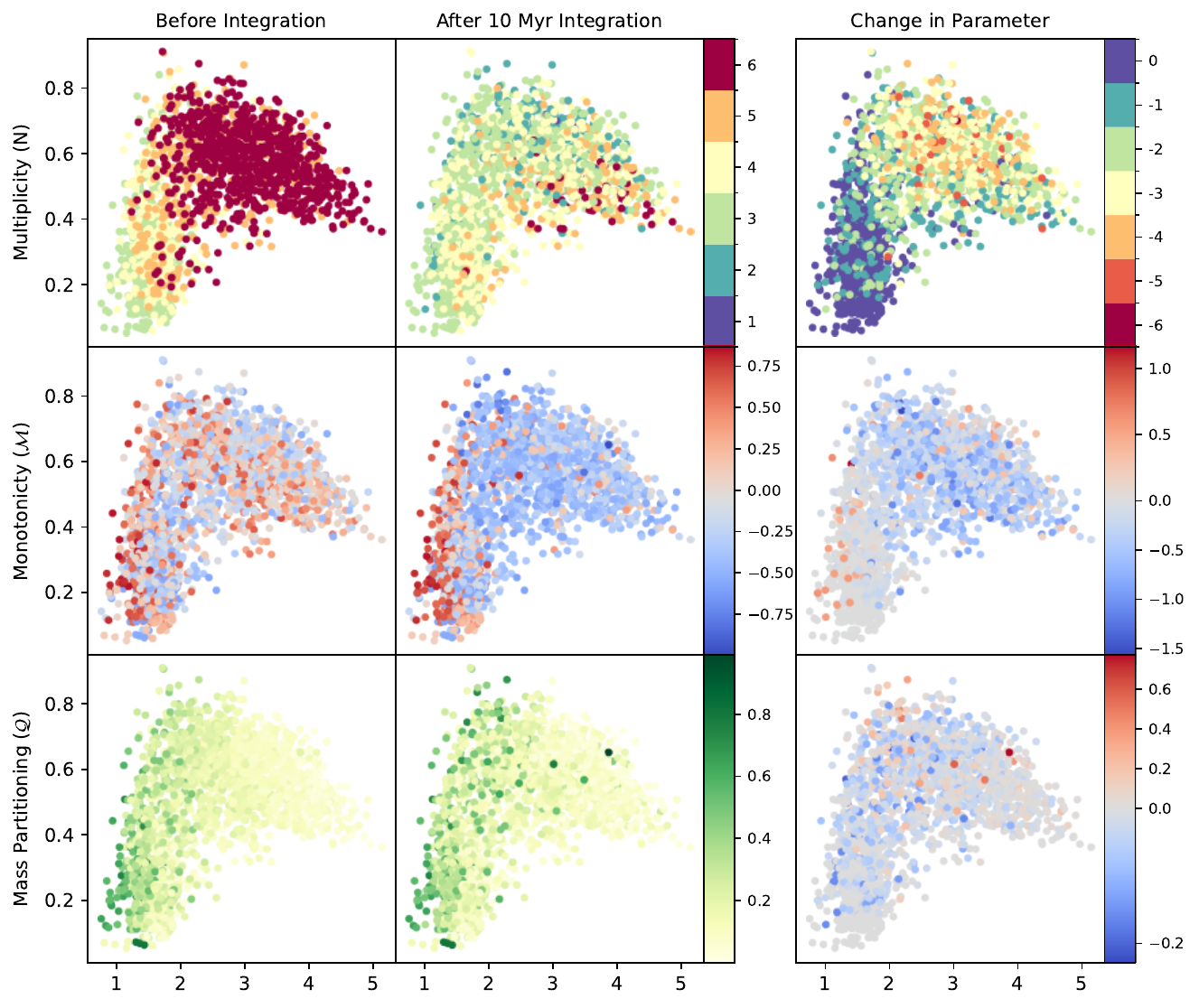}
    \caption{Scatter plots of ISO outcomes coloured by selected system architecture parameters, measured for all simulated systems. Left: initial values; Middle: final values; Right: net change, defined as the final minus initial value.}
    \label{fig:system1}
\end{sidewaysfigure}

\begin{sidewaysfigure}
    \centering
    \includegraphics{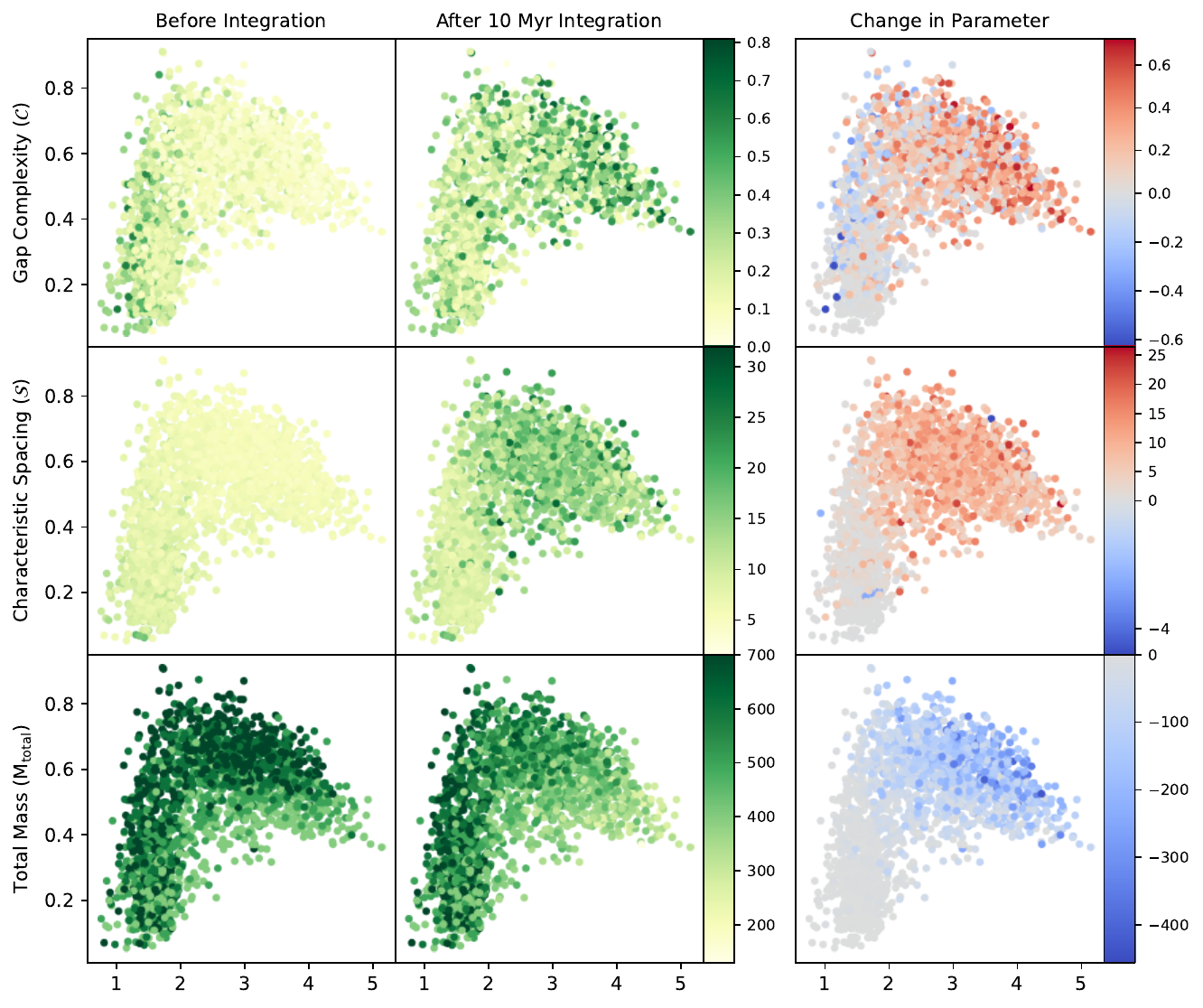}
    \caption{Same as \autoref{fig:system1}, for the remaining architecture parameters.}
    \label{fig:system2}
\end{sidewaysfigure}

To assess the information about system architecture that is contained in the ISO velocity distributions, we investigated correlations between ejection outcomes and planetary configuration metrics.
For this purpose, we adopted a slightly modified version of the exoplanet architecture classification scheme introduced by \citet{gilbert2020}.
Definitions of the architecture parameters are provided in \autoref{tab:system-params}, with the derivations and motivation discussed in detail by \citet{gilbert2020}.
\autoref{fig:system1} and \autoref{fig:system2} show the $v_\infty$–ejection fraction plane, coloured by system architecture parameters.
Each parameter is shown at three epochs: initial (left), final after 10 Myr (middle), and net change (right).
We then used $k$-means clustering to quantify whether the two lobes correspond to distinct system architectures.

\begin{deluxetable}{l l l l}
\tablecaption{Definitions of system architecture parameters, as used in this work. For full explanations, see \citet{gilbert2020}.\label{tab:system-params}}
\tablewidth{0pt}  
\renewcommand{\arraystretch}{1.2}
\tablehead{
\colhead{Parameter} & \colhead{Name} & \colhead{Definition} & \colhead{Meaning}
}
\startdata
$N$ & Multiplicity & $N$ & 
\parbox[t]{0.45\textwidth}{Number of planets in the system.} \\
$\mathcal{Q}$ & Mass Partitioning & $\left(\frac{N}{N-1}\right)\sum_{i=1}^{N}\left(m_i^{\ast}-\frac{1}{N}\right)^2$ & 
\parbox[t]{0.45\textwidth}{Ranges from 0 to 1. $\mathcal{Q}=0$ means equal masses; $\mathcal{Q}=1$ means all the mass is in one planet.} \\
$\mathcal{M}$ & Monotonicity & $\rho_s \mathcal{Q}^{1/N}$ & 
\parbox[t]{0.45\textwidth}{Ranges from $-1$ to $1$. Positive means mass increases with semi-major axis.} \\
$\mathcal{C}$ & Gap Complexity & $-K\sum_{i=1}^{n} p_i^{\ast}\log(p_i^{\ast})\left(p_i^{\ast}-\frac{1}{n}\right)^2$ & 
\parbox[t]{0.45\textwidth}{$\mathcal{C}=0$ for evenly spaced planets in log-period; $\mathcal{C}=1$ when one spacing dominates.} \\
$\mathcal{S}$ & Characteristic Spacing & $\mathrm{mean}(\Delta_H)$ & 
\parbox[t]{0.45\textwidth}{Average spacing in mutual Hill radii.} \\
$M$ & Total Mass & $\sum_{i=1}^{N} m_i$ & 
\parbox[t]{0.45\textwidth}{Total planetary mass.} \\
\enddata

\tablecomments{
[1] $m_i^{\ast} = \frac{m_i}{\sum_i^N m_i}$.  
[2] $p_i^{\ast} = \frac{\log(P'/P)}{\log(P_\mathrm{max}/P_\mathrm{min})}$.  
[3] $\Delta_H = \frac{a' - a}{r_H}$, with $r_H=\left(\frac{m'+m}{3}\right)^{1/3}\left(\frac{a'+a}{2}\right)$.
}

\end{deluxetable}

\subsection{Clustering}\label{sec:clusters}

\begin{figure}
    \centering
    \includegraphics{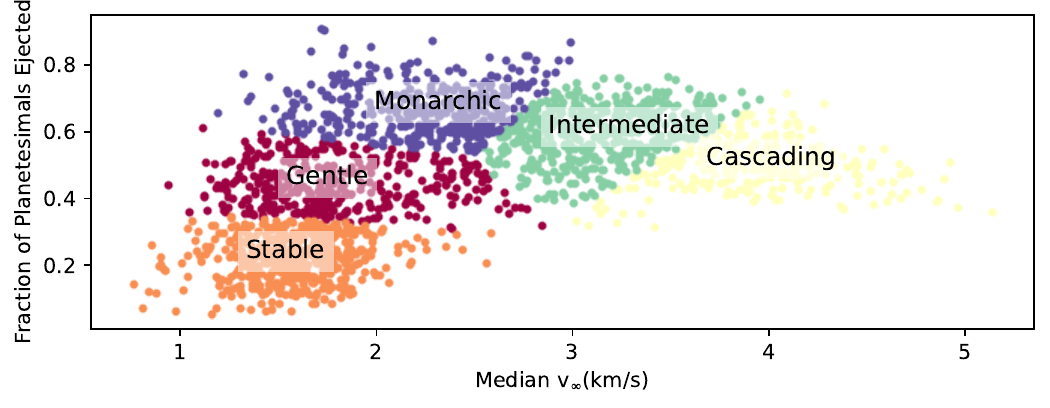}
    \caption{Separation of systems into 5 sub-clusters based on their location in the $v_\infty-f_\mathrm{ej}$ plane. The \textit{quiet evolution} cluster is essentially composed of the gentle and stable regions, and the \textit{catastrophic evolution} cluster is composed of the monarchic, intermediate and cascading sub-clusters.}
    \label{fig:clusters}
\end{figure}

The results of the $k$-means clustering are shown in \autoref{fig:clusters}, and the parameter distributions for each cluster are summarised in \autoref{tab:sim_results}. 
We first used $k=2$ to isolate the dominant two-mode structure in the $v_\infty-f_\mathrm{ej}$ plane, which has an ``L''-shaped morphology. 
In all cases, clustering was performed in the $v_\infty-f_\mathrm{ej}$ parameter space, where each dimension was standardised to have a mean of zero and variance of one. 
The clustering then used a Euclidean distance metric in this standardised space.
We assigned the descriptive labels \textit{catastrophic evolution} to the high-velocity horizontal branch and \textit{quiet evolution} to the low-velocity vertical branch.
To investigate what drives the transition from the green to the orange regions of the plot, we clustered with $k=5$, which essentially further subdivided the \textit{quiet evolution} cluster into two sub-clusters and the \textit{catastrophic evolution} branch into three additional sub-clusters.

\begin{table*}[t!]
\centering
\caption{%
Percentile summary of system architecture parameters for the two main clusters.
Values correspond to the 16th, 50th, and 84th percentiles of each population, at the beginning of the simulation (Initial), the end of 10 Myr of evolution (Final), and their difference, final minus initial (Change). 
The 16th and 84th percentile columns are de-emphasized for clarity.}
\label{tab:sim_results}
\renewcommand{\arraystretch}{1.15}
\setlength{\tabcolsep}{6pt}
\begin{tabular}{l *{3}{c} *{3}{c}}
\toprule
\multirow{2}{*}{\textbf{Parameter}} &
\multicolumn{3}{c}{\textbf{Quiet}} &
\multicolumn{3}{c}{\textbf{Catastrophic}} \\
\cmidrule(lr){2-4} \cmidrule(lr){5-7}
& 16\% & \textbf{50\%} & 84\% & 16\% & \textbf{50\%} & 84\% \\
\midrule
\multicolumn{7}{l}{\textbf{Summary}} \\[2pt]
Number of simulations & & \textbf{777} & & & \textbf{1184} & \\
$v_{\infty,\mathrm{med}}$ (km s$^{-1}$) & 1.345 & \textbf{1.611} & 1.871 & 2.205 & \textbf{2.927} & 3.754 \\
Fraction ejected & 0.171 & \textbf{0.284} & 0.430 & 0.485 & \textbf{0.593} & 0.688 \\
\midrule
\multicolumn{7}{l}{\textbf{Initial}} \\[2pt]
Multiplicity & 3 & \textbf{4} & 5 & 5 & \textbf{6} & 7\\
Mass Partitioning & 0.071 & \textbf{0.196} & 0.385 & 0.063 & \textbf{0.118} & 0.216 \\
Monotonicity & -0.420 & \textbf{-0.042} & 0.502 & -0.298 & \textbf{-0.019} & 0.378 \\
Gap Complexity & 0.130 & \textbf{0.219} & 0.368 & 0.057 & \textbf{0.104} & 0.186 \\
Characteristic Spacing & 6.775 & \textbf{8.246} & 9.931 & 5.156 & \textbf{5.946} & 6.916 \\
Total Mass (M$_\oplus$) & 400 & \textbf{500} & 700 & 400 & \textbf{600} & 700 \\
\midrule
\multicolumn{7}{l}{\textbf{Final}} \\[2pt]
Multiplicity & 3 & \textbf{3} & 4 & 3 & \textbf{3} & 4 \\
Mass Partitioning & 0.061 & \textbf{0.168} & 0.367 & 0.036 & \textbf{0.112} & 0.225 \\
Monotonicity & -0.453 & \textbf{-0.078} & 0.513 & -0.511 & \textbf{-0.356} & -0.090 \\
Gap Complexity & 0.128 & \textbf{0.225} & 0.369 & 0.050 & \textbf{0.261} & 0.496 \\
Characteristic Spacing & 7.605 & \textbf{9.101} & 11.361 & 10.603 & \textbf{14.279} & 19.057 \\
Total Mass (M$_\oplus$) & 400 & \textbf{500} & 692 & 351 & \textbf{449} & 570 \\
\midrule
\multicolumn{7}{l}{\textbf{Change}} \\[2pt]
Multiplicity & -1 & \textbf{0} & 0 & -3 & \textbf{-2} & -1 \\
Mass Partitioning & -0.040 & \textbf{0.000} & 0.000 & -0.055 & \textbf{0.003} & 0.047 \\
Monotonicity & -0.104 & \textbf{0.000} & 0.002 & -0.662 & \textbf{0.296} & 0.004 \\
Gap Complexity & -0.026 & \textbf{-0.000} & 0.021 & -0.080 & \textbf{0.1482} & 0.394 \\
Characteristic Spacing & -0.024 & \textbf{0.028} & 2.648 & 4.233 & \textbf{8.237} & 13.187 \\
Total Mass (M$_\oplus$) & -10.6 & \textbf{0.000} & 0.000 & -188.0 & \textbf{-89.4} & -31.0 \\
\bottomrule
\end{tabular}
\end{table*}

\subsection{Cluster System Architecture Trends}\label{trends}

Systems in the catastrophic evolution cluster are marked by violent dynamical activity, ejecting many high-speed ISOs and often losing several planets.
By contrast, quiet evolution systems remain comparatively calm, showing smaller orbital rearrangements and ejecting fewer planets.

The most significant system architectural features of catastrophic systems are large increases in characteristic spacing, with planets spreading apart by an average of 8.237 mutual Hill radii.
These systems also experience substantial mass loss, with a median loss of 89.4 M$_\oplus$.
They undergo significant reductions in multiplicity, losing a median of 2 planets per system, confirming that planet ejection is a common occurrence in these systems.
Monotonicity also typically increases ($\Delta \mathcal{M} = 0.296$), consistent with preferential removal of lower-mass planets or inward scattering.

The quiet systems, in contrast, retain much of their initial architecture.
They lose very little mass (most systems ejecting no planets), and experience only minor changes in characteristic spacing ($0.028$ mutual Hill radii).
They begin with higher gap complexities, with a median of 0.219.
Wider initial spacing allows more irregular planet placement during initialization, increasing gap complexity.
Quiet systems also show higher initial mass partitioning, 0.196,
(catastrophic systems, $\mathcal{Q}= 0.118$), meaning they are hosting planets with more asymmetric masses.
Lower multiplicities make asymmetric mass distributions more likely, due to the $\alpha=1$ parametrisation of the Dirichlet distribution.
The mass asymmetry in quiet systems likely suppresses chaotic multi-body interactions, providing a stabilizing influence.
Most quiet systems preserve their initial monotonicity, indicating limited planetary reordering and ejections.

Both clusters converge to a median final multiplicity of three planets and have similar final total masses (500 M$\oplus$ and 449 M$\oplus$).
This configuration appears to be a dynamically stable outcome for the high-mass architectures explored here.
Within the broader catastrophic and quiet groups, we further explored dividing the simulations into finer structure: three catastrophic sub-clusters and two quiet sub-clusters.
Although this five-way division was guided by the elbow method for $k$-means clustering, the choice is still somewhat arbitrary.
Our goal in subdividing the parameter space was to identify architectural commonalities within each region, rather than to claim that the data naturally separate into five intrinsic classes.
These classes may not translate to systems with different mass–multiplicity regimes.
The average system-architecture parameters for each sub-cluster are listed in \autoref{tab:subclusters}.
Together, these subdivisions reveal a continuum of dynamical pathways and show how planetary architecture modulates ISO production.

\begin{table*}[t!]
\centering
\caption{%
Percentile summary of system architecture parameters for the five sub-clusters.
Values correspond to the 16th, 50th, and 84th percentiles of each population
at the beginning of the simulation (Initial), the end of 10 Myr of evolution (Final),
and their difference, final minus initial (Change). The 16th and 84th percentile columns are de-emphasised for clarity.}
\label{tab:subclusters}

\renewcommand{\arraystretch}{1.1}
\setlength{\tabcolsep}{2pt}

\begin{tabular}{lccccccccccccccc}
\hline
\multirow{2}{*}{\textbf{Parameter}}     & \multicolumn{3}{c|}{\textbf{Stable}}                & \multicolumn{3}{c|}{\textbf{Gentle}}                & \multicolumn{3}{c|}{\textbf{Monarchic}}             & \multicolumn{3}{c|}{\textbf{Intermediate}}          & \multicolumn{3}{c}{\textbf{Cascading}} \\ \cline{2-16} 
                                        & 16\%  & \textbf{50\%}  & \multicolumn{1}{c|}{84\%}  & 16\%  & \textbf{50\%}  & \multicolumn{1}{c|}{84\%}  & 16\%  & \textbf{50\%}  & \multicolumn{1}{c|}{84\%}  & 16\%  & \textbf{50\%}  & \multicolumn{1}{c|}{84\%}  & 16\%     & \textbf{50\%}     & 84\%    \\ \hline
\multicolumn{16}{l}{\textbf{Summary}}                                                                                                                                                                                                                                                                    \\
Number of Simulations                   &       & \textbf{491}   & \multicolumn{1}{c|}{}      &       & \textbf{358}   & \multicolumn{1}{c|}{}      &       & \textbf{386}   & \multicolumn{1}{c|}{}      &       & \textbf{457}   & \multicolumn{1}{c|}{}      &          & \textbf{269}      &         \\
$v_{\infty,\mathrm{med}}$ (km s$^{-1}$) & 1.320 & \textbf{1.594} & \multicolumn{1}{c|}{1.837} & 1.411 & \textbf{1.737} & \multicolumn{1}{c|}{2.250} & 1.833 & \textbf{2.293} & \multicolumn{1}{c|}{2.578} & 2.812 & \textbf{3.092} & \multicolumn{1}{c|}{3.454} & 3.551    & \textbf{3.902}    & 4.290   \\
$f_{\mathrm{ej}}$                       & 0.146 & \textbf{0.227} & \multicolumn{1}{c|}{0.295} & 0.368 & \textbf{0.435} & \multicolumn{1}{c|}{0.518} & 0.595 & \textbf{0.659} & \multicolumn{1}{c|}{0.741} & 0.514 & \textbf{0.597} & \multicolumn{1}{c|}{0.672} & 0.426    & \textbf{0.497}    & 0.565   \\ \hline
\multicolumn{16}{l}{\textbf{Initial}}                                                                                                                                                                                                                                                                    \\
Multiplicity                            & 3     & \textbf{3}     & \multicolumn{1}{c|}{4}     & 3     & \textbf{4}     & \multicolumn{1}{c|}{6}     & 4     & \textbf{5}     & \multicolumn{1}{c|}{7}     & 5     & \textbf{6}     & \multicolumn{1}{c|}{7}     & 5        & \textbf{6}        & 7       \\
Mass Partitioning                       & 0.07  & \textbf{0.20}  & \multicolumn{1}{c|}{0.43}  & 0.08  & \textbf{0.18}  & \multicolumn{1}{c|}{0.33}  & 0.13  & \textbf{0.19}  & \multicolumn{1}{c|}{0.28}  & 0.07  & \textbf{0.10}  & \multicolumn{1}{c|}{0.16}  & 0.03     & \textbf{0.07}     & 0.11    \\
Monotonicity                            & -0.43 & \textbf{-0.03} & \multicolumn{1}{c|}{0.52}  & -0.38 & \textbf{-0.09} & \multicolumn{1}{c|}{0.39}  & -0.29 & \textbf{0.00}  & \multicolumn{1}{c|}{0.50}  & -0.33 & \textbf{-0.06} & \multicolumn{1}{c|}{0.36}  & -0.22    & \textbf{0.03}     & 0.33    \\
Gap Complexity                          & 0.13  & \textbf{0.23}  & \multicolumn{1}{c|}{0.38}  & 0.10  & \textbf{0.19}  & \multicolumn{1}{c|}{0.32}  & 0.07  & \textbf{0.13}  & \multicolumn{1}{c|}{0.24}  & 0.06  & \textbf{0.10}  & \multicolumn{1}{c|}{0.18}  & 0.05     & \textbf{0.08}     & 0.14    \\
Characteristic Spacing                  & 7.07  & \textbf{8.60}  & \multicolumn{1}{c|}{10.29} & 6.17  & \textbf{7.46}  & \multicolumn{1}{c|}{8.99}  & 5.19  & \textbf{6.12}  & \multicolumn{1}{c|}{7.03}  & 5.06  & \textbf{5.82}  & \multicolumn{1}{c|}{6.79}  & 5.23     & \textbf{5.87}     & 6.70    \\
Total Mass (M$_\oplus$)                 & 400   & \textbf{500}   & \multicolumn{1}{c|}{600}   & 400   & \textbf{600}   & \multicolumn{1}{c|}{700}   & 500   & \textbf{600}   & \multicolumn{1}{c|}{700}   & 400   & \textbf{600}   & \multicolumn{1}{c|}{700}   & 400      & \textbf{500}      & 600     \\ \hline
\multicolumn{16}{l}{\textbf{Final}}                                                                                                                                                                                                                                                                      \\
Multiplicity                            & 3     & \textbf{3}     & \multicolumn{1}{c|}{4}     & 3     & \textbf{3}     & \multicolumn{1}{c|}{4}     & 2     & \textbf{3}     & \multicolumn{1}{c|}{4}     & 2     & \textbf{3}     & \multicolumn{1}{c|}{4}     & 3        & \textbf{4}        & 5       \\
Mass Partitioning                       & 0.07  & \textbf{0.18}  & \multicolumn{1}{c|}{0.38}  & 0.05  & \textbf{0.16}  & \multicolumn{1}{c|}{0.30}  & 0.08  & \textbf{0.18}  & \multicolumn{1}{c|}{0.35}  & 0.03  & \textbf{0.09}  & \multicolumn{1}{c|}{0.18}  & 0.03     & \textbf{0.07}     & 0.13    \\
Monotonicity                            & -0.46 & \textbf{-0.05} & \multicolumn{1}{c|}{0.53}  & -0.47 & \textbf{-0.24} & \multicolumn{1}{c|}{0.34}  & -0.53 & \textbf{-0.37} & \multicolumn{1}{c|}{0.05}  & -0.50 & \textbf{-0.36} & \multicolumn{1}{c|}{-0.13} & -0.48    & \textbf{-0.32}    & -0.07   \\
Gap Complexity                          & 0.13  & \textbf{0.22}  & \multicolumn{1}{c|}{0.37}  & 0.12  & \textbf{0.22}  & \multicolumn{1}{c|}{0.35}  & 0.00  & \textbf{0.23}  & \multicolumn{1}{c|}{0.43}  & 0.00  & \textbf{0.28}  & \multicolumn{1}{c|}{0.51}  & 0.12     & \textbf{0.33}     & 0.55    \\
Characteristic Spacing                  & 7.60  & \textbf{8.98}  & \multicolumn{1}{c|}{11.00} & 7.84  & \textbf{9.60}  & \multicolumn{1}{c|}{13.63} & 10.46 & \textbf{14.35} & \multicolumn{1}{c|}{18.45} & 11.57 & \textbf{15.19} & \multicolumn{1}{c|}{19.45} & 9.85     & \textbf{13.37}    & 18.61   \\
Total Mass (M$_\oplus$)                 & 400   & \textbf{500}   & \multicolumn{1}{c|}{600}   & 400   & \textbf{528}   & \multicolumn{1}{c|}{692}   & 430   & \textbf{528}   & \multicolumn{1}{c|}{632}   & 357   & \textbf{443}   & \multicolumn{1}{c|}{532}   & 294      & \textbf{363}      & 420     \\ \hline
\multicolumn{16}{l}{\textbf{Change}}                                                                                                                                                                                                                                                                     \\
Multiplicity                            & -1     & \textbf{0}     & \multicolumn{1}{c|}{0}     & -2     & \textbf{-1}     & \multicolumn{1}{c|}{0}     & -3     & \textbf{-2}     & \multicolumn{1}{c|}{-1}     & -4     & \textbf{-2}     & \multicolumn{1}{c|}{-1}     & -3        & \textbf{-2}        & -1       \\
Mass Partitioning                       & -0.02  & \textbf{0.00}  & \multicolumn{1}{c|}{0.00}  & -0.07  & \textbf{0.00}  & \multicolumn{1}{c|}{0.00}  & -0.07 & \textbf{0.00}  & \multicolumn{1}{c|}{0.07}  & -0.05 & \textbf{0.01}  & \multicolumn{1}{c|}{0.04}  & -0.03 & \textbf{0.00}  & 0.03    \\
Monotonicity                            & 0.00  & \textbf{0.00}  & \multicolumn{1}{c|}{0.00}  & -0.01  & \textbf{0.00}  & \multicolumn{1}{c|}{0.00}  & -0.65  & \textbf{-0.29}  & \multicolumn{1}{c|}{-0.03}  & -0.69 & \textbf{-0.29}  & \multicolumn{1}{c|}{-0.05}  & -0.65    & \textbf{-0.34}     & -0.03    \\
Gap Complexity                          & -0.01 & \textbf{0.00}  & \multicolumn{1}{c|}{0.01}  & -0.09 & \textbf{0.00}  & \multicolumn{1}{c|}{0.15}  & -0.11 & \textbf{-0.08} & \multicolumn{1}{c|}{0.31}  & -0.06 & \textbf{-0.17} & \multicolumn{1}{c|}{0.40}  & 0.00    & \textbf{0.25}    & 0.47 \\
Characteristic Spacing                  & -0.04 & \textbf{-0.01} & \multicolumn{1}{c|}{1.29}  & 0.00 & \textbf{-1.73} & \multicolumn{1}{c|}{6.67}  & 3.65 & \textbf{8.28} & \multicolumn{1}{c|}{12.4} & 5.45 & \textbf{9.33} & \multicolumn{1}{c|}{13.9} & 3.81   & \textbf{7.52}    & 12.78   \\
Total Mass (M$_\oplus$)                 & -2.0     & \textbf{0.0}     & \multicolumn{1}{c|}{0.0}     & -43.3     & \textbf{1.45}     & \multicolumn{1}{c|}{0.0}    & -143.5    & \textbf{-76.7}    & \multicolumn{1}{c|}{-26.0}   & -211.3    & \textbf{-106.3}   & \multicolumn{1}{c|}{-43.5}   & -215.9       & \textbf{-108.2}      & -32.9     \\ \hline
\end{tabular}
\end{table*}

\subsubsection{Quiet Sub-Clusters}

The quiet evolution cluster was divided into two sub-populations, separated at an ejection fraction of $\sim0.3$.  
Although both sub-clusters exhibit similar median $v_\infty$ values ($1.594$ and $1.737$ km s$^{-1}$), their ejection fractions differ substantially (0.227 to 0.435), motivating this distinction.

We refer to the lower-ejection fraction systems as \textit{stable systems}. 
These systems are not entirely static, but their architecture parameters change far less than in other clusters. 
For every parameter, the mean change is near zero.
These systems have the highest initial gap complexity ($0.23$), which remains nearly constant, and the largest characteristic spacings ($8.60$ initially, increasing to $8.98$ mutual Hill radii). 
This indicates that these systems are sparsely populated with planets initially, and unlikely to experience dynamical instabilities.
The monotonicity also shows negligible change, indicating no reordering of planetary masses.  
These properties suggest that stable systems do not undergo large-scale orbital changes.  
They also begin with low multiplicities (median 3) and have a median of zero planetary ejections, reinforcing the picture of long-lived, sparsely-populated planetary architectures. 
The typical ejected bodies from these systems are small, with the upper $84\%$ boundary at a mass loss of 2 M$_\oplus$, and most commonly no planets are ejected.
As a result, these systems eject few ISOs, and those that do escape are likely removed through gentle unbinding or single scattering events, which can only experience a finite velocity kick of $v_\mathrm{esc}/\sqrt{2}$ \citep{wyatt2017}.  

The second sub-cluster, which we term the \textit{gentle ejectors}, exhibits more dynamical activity while still avoiding catastrophic disruption.  
Compared to the stable systems, these have slightly higher initial multiplicities (4) and starting masses ($600$ M$_\oplus$), and lose on average $1$ planet over the integration.  
Their characteristic spacings grow from a median $7.46$ to $9.6$, larger than in the stable systems but far below the final spacings of $\sim 14$ attained by catastrophic systems.  
This indicates that the orbits of planets in gentle ejector systems expand outward, but remain relatively tightly packed at the end of the evolution.  
Gap complexity and monotonicity changes are minimal, reinforcing the picture of limited reordering.  
Their higher median starting masses ($600$ M$_\oplus$) and relatively small mass loss (1 M$_\oplus$) lead to these systems retaining the largest amount of mass (although strongly overlapping with the monarchic ejector sub-cluster).  
We interpret this as evidence that most planets remain bound, and that these systems produce ISOs largely through gradual orbital evolution rather than direct ejection by planets. 
Planet–planetesimal interactions slowly modify small-body orbits, lifting them onto unstable trajectories over long timescales. 
The ISO flux in the time-frame we examine is therefore driven less by impulsive scattering events and more by cumulative orbital changes. 
This highlights that significant ISO production within 10 Myr does not necessarily require strong, violent instabilities, but can also emerge from comparatively quiescent dynamical pathways.
Both of the quiet clusters release ISOs at similar velocities, so within the 10 Myr timescale the more massive gentle ejector systems are able to eject more material due to their higher starting masses, but through a similar process as the stable systems. 

\subsubsection{Catastrophic Sub-Clusters}

For the $k=5$ clustering, the catastrophic evolution cluster has essentially been separated into three distinct sub-clusters, which we term \textit{monarchic ejectors}, \textit{intermediate ejectors}, and \textit{cascading ejectors} (\autoref{fig:clusters}).

The monarchic ejector systems are characterized by the highest ejection fractions of any cluster ($0.659$) but also the lowest median ejection velocity among the catastrophic groups ($v_{\infty,\mathrm{med}} = 2.293$ km s$^{-1}$). 
The architectural parameters of these systems have distinct trends from the other clusters, allowing us to hypothesise the driving factors behind these differences.
These systems begin with the highest average total masses ($600$ M$_\oplus$) but the lowest median initial multiplicities ($5$ planets), and they lose an average of $2$ planets over the course of the simulation, the lowest for any of the catastrophic sub-clusters.
They also exhibit the highest initial gap complexity ($0.13$) and characteristic spacing (6.12) within the catastrophic sub-clusters.
Importantly, they maintain relatively high mass partitioning (initial $0.19$, final $0.18$), a value closer to the quiet systems (0.18) than to other catastrophic sub-clusters.
This suggests that a single massive planet retains gravitational dominance throughout the simulation.
These properties suggest that the dynamical instability in the system is primarily driven by one body -- the ``monarch'' -- which dominates scattering and ejects a large fraction of the planetesimals. 
Ejections in this regime are therefore efficient but comparatively orderly, controlled by the gravitational influence of a single dominant planet.
Minimal planetary orbital architectural changes occur, and the planetesimals are thus ejected in the same manner as they were by the quiet systems, but with a much more massive driver. 
This shows the continuum between our quiet ejector systems and the clusters which eject faster ISOs.

The cascading ejector systems display the most dramatic architectural evolution.
They start with the highest multiplicities ($6$ planets) and lowest initial gap complexities ($0.08$) and characteristic spacings (5.87), reflecting tightly packed, quasi–``peas-in-a-pod'' architectures.
They have the lowest mass partitioning (0.07), meaning all the planets in these systems are similar in size. 
The cascading systems also start with the lowest mass of all of the catastrophic clusters (500 M$_\oplus$), and lose the most mass throughout the simulation duration (108 M$_\oplus$), the remnants of these systems being the lowest mass of all the systems, while still retaining the highest multiplicity (median 4, 84th percentile 5).
The minimal changes to the mass partitioning indicate that the ejected planets are a similar mass to the remaining planets.
The instabilities in these systems are thus hypothesised to take the pathway of chain reactions: once one planet is perturbed, tightly packed neighbours are destabilized, triggering further ejections and reordering.
Planets cannot efficiently eject each other, but orbital perturbations can lead to planets and planetesimals kicked onto eccentric orbits.
This creates path-crossing orbits that lead to frequent close encounters, and violent ejection events with high ejection velocities.
The result is a ``cascade'' of dynamical interactions, with large-scale disruption of the planetary system and intense stirring of the planetesimal disc.

Overall, these sub-clusters illustrate that there are two pathways to ejecting many ISOs.
Systems can either have a single massive planet which is able to drive slow ejections, or tightly-packed, less massive planets that can directly influence the planetesimal orbits when perturbing each other. 
The intermediate ejector systems occupy a middle ground, with ejection fractions of $0.597$ and median ejection velocities around $3.092$ km s$^{-1}$ (between the monarchic and cascading groups).
They begin with initial masses, multiplicity, mass partitioning, gap complexity, and characteristic spacing also all intermediary.
These systems fill in the continuum between the ejection mechanisms of the two types of catastrophic systems, and seem to consist of systems with less extreme versions of the system architectures in the other two clusters.
Taken together, this points to instabilities driven by many intermediate-mass planets rather than a single dominant one.
The resulting scattering has a smaller gravitational influence than the monarchic systems, so they cannot eject quite as many planetesimals, but do not have the similarly-sized planets which make the cascading ejectors so susceptible to chain events.

\section{Discussion}
\label{sec:discussion}

\subsection{Dynamical Instability as an Interstellar Object Production Mechanism}

Many dynamical instability triggers are thought to occur early in systems' lifetimes, when there are the most planetesimals available to eject.
High-resolution simulations of the Solar System, initialized with realistic conditions, demonstrate that a very large fraction of planetesimals are eventually ejected, with \citetalias{nesvorny2023} finding that $\sim$93\% of the original small-body population is lost to interstellar space.
We show here that 0.2-0.9 of the planetesimal disks we initiate are also ejected, becoming ISOs, in just the first 10 Myr of the evolution.
If dynamical instability is as ubiquitous as has been suggested, then this process is indeed the dominant ISO production mechanism across the Galaxy.

Ejection velocity distributions evolve with system age.
Planetesimals must first be scattered onto wide, high-perihelion orbits before the slowest velocity ejections can occur.
The \citetalias{nesvorny2023} simulations illustrate how the source regions of ISOs shift as a function of time, with distant ejections becoming increasingly common at later stages (see \autoref{fig:perihelion-scatter}).
This temporal evolution may also influence the expected composition of ISOs: rocky bodies are preferentially ejected shortly after gas dispersal, while icy material is more likely to dominate the later, slower-ejection population.
As shown in \autoref{fig:semi-ax-vinf} and in the velocity distribution snapshots of \autoref{fig:all-sims-hist}, the fastest ISOs are typically produced early in a system’s lifetime, whereas the slower, low-velocity tail emerges much later. 
In fact, the lowest-velocity ejections ($v \lesssim 2\times 10^{-2}$ km s$^{-1}$) in \citetalias{nesvorny2023} only appear after $\sim$20 Myr of evolution.

The physical mechanisms responsible for these two regimes differ. 
Fast ejections generally result from a single strong scattering with a planet, occurring relatively close to the host star (see green points in \autoref{fig:time-series-individual}). 
In contrast, slow ejections are usually preceded by a gradual outward diffusion in semi-major axis before eventual unbinding, as seen in \autoref{fig:semi-ax-vinf}. 
This highlights that dynamical instability produces a spectrum of outcomes: early, violent scattering yields fast ISOs, while orbital diffusion leads to the delayed production of slower-moving ISOs.

\subsection{Velocity Distributions as Tracers of System Architecture}

\begin{figure}
    \centering
    \includegraphics{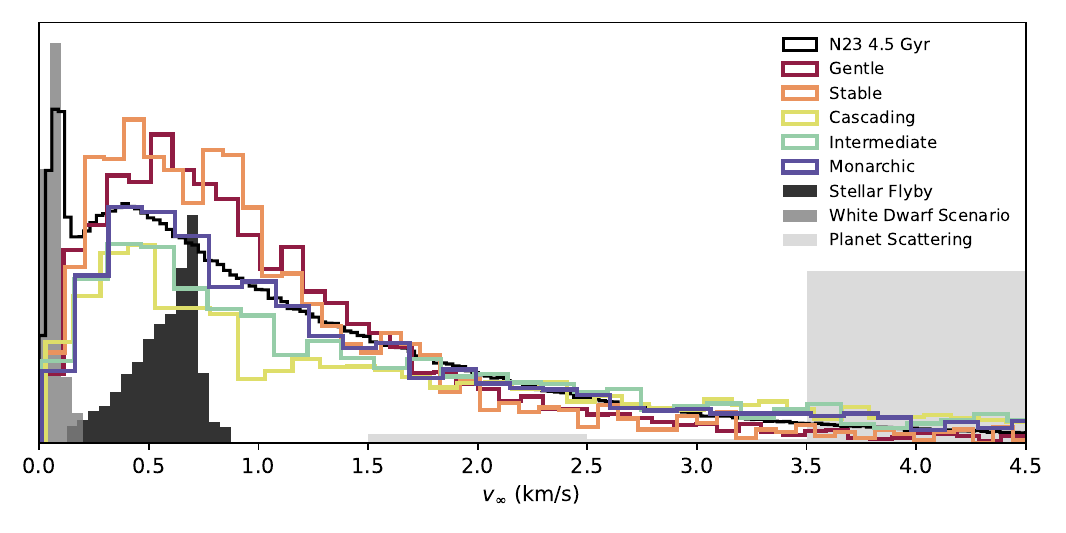}
    \caption{ISO ejection velocity distributions for various mechanisms.
    This is a modified version of the plot from \cite{Pfalzner2021}, with an example of a velocity distribution from a stellar flyby from that paper, white dwarf scenario data from unpublished supplementary material of \cite{veras2014}, and planet–planet scattering data from \cite{adams2005}. 
    To these literature datasets, we add all ejections from the \citetalias{nesvorny2023} simulation (black line), and representative example distributions from our simulation clusters (\autoref{fig:clusters}). For each cluster, the simulation with the architecture parameters closest to the reported medians in \autoref{tab:subclusters} is displayed.
    Note that the \cite{adams2005} planet–planet scattering simulations had a higher velocity range, and most of the histogram has been truncated here. 
    Data from both the planet-planet scattering simulations and the stellar flyby simulations are ambiguous as to whether the reported velocities are actually $v_\infty$ or some other related velocity. 
    Following \citet{Pfalzner2021}, we assume they are $v_\infty$ values for the purposes of this plot.}
    \label{fig:ejection_mechs}
\end{figure}

To compare ISO velocity distributions across different dynamical ejection mechanisms, in \autoref{fig:ejection_mechs} we adapt a compilation figure from \cite{Pfalzner2021}, incorporating literature to date and our \autoref{fig:clusters} cluster velocity medians.
%In \autoref{fig:ejection_mechs}, we overlay our results with those from \citetalias{nesvorny2023} and with the five-planet, 700 M$_\oplus$ simulation presented in \autoref{sec:individual-system}.
%Excluding the low-velocity peak, the \citetalias{nesvorny2023} distribution resembles our results (in linear space), and 
Ejection velocities vary significantly between ISO ejection mechanisms.
\citet{veras2014}'s white dwarf remnant scenario produces the slowest ISOs, with mean velocities of only $\sim$0.065 km s$^{-1}$, while stellar flybys yield moderately higher values ($\sim$0.56 km s$^{-1}$). 
The subset of planet–planet scattering explored by \citet{adams2005}, by contrast, appear to generate much faster\footnote{The large velocities reported in \citet{adams2005} may not be $v_\infty$, but rather a velocity when the ejected object is still close to the planet.} ISOs, with a median of 5.4 km s$^{-1}$. 
For comparison, \citetalias{nesvorny2023}'s Solar System simulations report a median of 0.96 km s$^{-1}$.
%ur wide parameter space instability simulations give 0.91 km s$^{-1}$. 
The low-velocity component in \citetalias{nesvorny2023} aligns well with the peak of the white dwarf–remnant scenario.
This may reflect the fact that late-time, low-energy ejections more closely resemble the stripping of a diffuse, loosely bound reservoir of material rather than the direct scattering of planetesimals by planets. 
Both the \citetalias{nesvorny2023} and \citet{veras2014} simulations included stellar flybys and Galactic tides, which could also contribute to the formation of this extended low-velocity tail.

Dynamical instabilities are distinctive producers of ISOs in that they span nearly the entire accessible ISO velocity space. 
This breadth arises from the variety of gravitational encounters possible, ranging from violent, close-in scatterings that launch material at high speed to more gradual, long-term orbital diffusion that produces slow, low-energy unbindings. 
Even within our relatively coarse, high-mass parameter space, we recover meaningful substructure in the velocity distributions.
Given the enormous diversity of known exoplanet architectures, there is still commonality around instability-generated distributions relative to other production mechanisms. 
The ISO velocity distributions arising from planetary systems across the Galaxy are equally diverse.

If we could measure an ISO's ejection velocity, we could infer aspects of its birth system's architecture.
This is a difficult task due to the influence of the Galaxy and its inhomogeneities (see section \ref{sec:gal}), but is within the realm of possibility for very young ISOs (ages $\lesssim$ Myr).
These may be able to be traced, along with a large number of stars in the local volume observed by \textit{Gaia}, back in time to identify the progenitor. 
The chances of this technique succeeding are very low \citep{Forbes2024}, especially for multi-Gyr-old ISOs like 3I/ATLAS \citep{Hopkins2025b, Taylor2025}.
This has been attempted unsuccessfully for all three ISOs detected thus far within the Solar System \citep[e.g.][for 3I]{Perez-Couto2025}.
A more broadly viable route would be to use our results as an input in a Galactic-scale forward model for the distribution of ISO velocities.

%Despite these systematic trends, using velocity distributions to identify the origin of ISOs remains effectively impossible. 
%Once ejected, ISOs inherit additional velocity shifts from the relative motion of their host system with respect to the Sun, and their orbits are further dispersed by long-term perturbations such as the galactic tide \citep[see][]{Forbes2024}. 
%These effects erase much of the diagnostic information carried by the initial distribution.
%Compounding this, velocity distributions evolve internally: immediately after an instability, systems eject fast ISOs with relatively narrow dispersions, but as instabilities continue the mean velocity decreases and the spread broadens. 
%Consequently, even if a coherent stream of ISOs could be observed, its present-day kinematics would be too diluted and time-dependent to reliably reconstruct the properties of the progenitor system.

\subsection{Connection to Observations and Galactic Population}
\label{sec:gal}

For a rare handful of locally passing ISOs, velocity is not the only diagnostic: observational campaigns can provide detailed constraints on their chemistry.
Inbound measurements of the coma composition of the most recent interstellar comet 3I/ATLAS \citep[e.g.][]{opitom2025, puzia2025, yang2025, cordiner2025} have found a range of volatiles. 
Coma molecule production rates were also measured throughout perihelion passage for 2I/Borisov \citep[e.g.][]{Deam2025}. 
Simulations suggest that future ISOs discovered by the Vera C. Rubin Observatory will also spend several hundred days at bright enough magnitudes to allow spectroscopy \citep{Dorsey2025}.
While we do not explore chemistry here, understanding the compositional differences between ISOs released in the different ways illustrated in \autoref{fig:ejection_mechs} could allow for better linking of observed ISOs to the properties of their progenitor systems. 
Some work has been done by \citet{hopkins2023} linking the metallicity gradient of the Galaxy to the ISO population, however using the composition of observed ISOs may provide more information about their release mechanism.

While close encounters with their planets can provide planetesimals with a larger kick, we have found that the typical ejection velocities of ISOs are much slower than the maximum limit.
This is true even in systems where planets more massive than Jupiter are present, and in systems where the planets undergo a dynamical instability. 
This is because the typical interaction is much more modest, and the planetesimals execute a random walk in energy space until they are eventually unbound, as is expected even in single-planet scattering \citep[e.g.][]{Hadden2024,Huang2025}.

Comparatively slow ejection velocities have substantial downstream effects for how the planetesimals behave in the Galactic potential. 
When the planetesimals leave their home star system, they spread out into a long thin tidal stream due to the initial spread in velocities and the differential rotation implied by the Galaxy's rotation curve. 
The typical length, width, and height of the streams are each proportional to the internal velocity dispersion of the stream \citep{Dehnen2018,Carlberg2024,Forbes2024}, which in turn is initially set by the ejection velocity distribution. 
The streams will experience subsequent heating and perturbations in the Galactic potential, but will keep the imprint of the initial velocity distribution. 
The low velocities we find here imply that the streams will be denser, and therefore that fewer streams will contain the Sun \citep{Forbes2024} at a given overall rate that ISOs encounter the Solar System. 
As a result, the rate of ISOs we will see in sky surveys may be supplied by a smaller number of streams, meaning that we are more likely to see multiple ISOs from the same stream star, and that the ISOs we see will be younger and more feasible to trace back to their parent star.

The velocity of the interstellar objects relative to their parent star is initially far smaller than the typical spread in velocities of the stellar population, which is about 10-50 km s$^{-1}$ depending on the component of motion and the age of the stars under consideration \citep[e.g.][]{Holmberg2009,Gaia2018}. 
This validates the frequent earlier assumption that the ISO population may be traced by the population of stars.
Note that this still requires appropriate reweighting of stars to account for trends in the ISO production rate as a function of other stellar properties, most notably metallicity \citep{Hopkins2025}. 
Eventually the ISOs will also be subject to dynamical heating due to perturbers in the Galactic potential \citep[e.g.][]{spitzer1951,lacey1984,forbes2012} on timescales $\gtrsim 100$ Myr. 
This increases the spread in velocities of ISOs at a given location.
Meanwhile, even without heating, the velocity of an ISO will typically be substantially different from its particular host star because they are on different orbits and therefore drift apart in the Galaxy. A given stream of ISOs will have a length of order its velocity dispersion times its age \citep{Forbes2024}.
Nonetheless, because the stars are subject to the same dynamical heating and orbit in the same potential, once any biases\footnote{Admittedly these biases are not known, but trends with the occurrence of planets as a function of host star properties may provide guidance \citep[see][for a review]{Zhu2021}.} in ISO production as a function of stellar properties have been accounted for, the only remaining difference originates from the ejection velocities. 
Because the ejection velocities are small (\autoref{fig:ejection_mechs}) compared to the typical spread in stellar velocities, it is the dynamics of the Galaxy and not the ejections, that determine the distribution of velocities of the ISOs we will observe in the Solar System. 
Stream-related effects, namely the possibility of finding multiple ISOs from the same star, are much more sensitive to the ejection velocity distribution \citep{Forbes2024}.

\subsection{Re-scaleability for wider use}

Our simulations are described as planetary systems orbiting a Solar-mass, with orbital distances expressed in au and masses in Solar masses.
However, because gravity is scale-free and our simulations omit scale-dependent effects such as the Galactic tide, the results can be rescaled, to represent systems of different stellar masses or orbital scales.
If we scale the mass by a factor $x$, so that the new masses in the system are $\tilde{M} = xM$, and similarly scale the distances by a factor $y$, then the ejection fractions and all of the system architecture parameters, besides Total Mass, will be unchanged, but the ejection velocities will be affected. 
In particular, all velocities in the simulation, including ejection velocities, will be scaled as 
\begin{equation}
    \tilde{v} = v \sqrt{\frac{x}{y}}.
\end{equation}
Thus, more extended or lower-mass systems will eject material more slowly, while higher-mass or more compact systems will produce faster ejecta in physical units (km s$^{-1}$).

It is important to note that we do not have complete freedom to set the ejection velocities to arbitrary values by carrying out this rescaling. 
Stellar and disc properties impose correlations between $x$ and $y$.
For example, lower-mass stars are generally associated with smaller protoplanetary discs \citep[e.g.][]{Andrews2020}, which naturally lead to more compact planetary systems.
This coupling implies that realistic rescaling will shift the velocity distributions only modestly, rather than producing drastic changes.

\section{Conclusion}

In this work, we have investigated how dynamical instabilities influence the ejection of interstellar objects (ISOs) from planetary systems with massive planets and planetesimal discs.
By simulating 2,461 systems for 10 Myr, and exploring a wide range of total system masses (300-800 M$_\oplus$), multiplicities (3-7 planets), and initial orbital configurations, we characterise the velocity and inclination distributions of ejected material as a function of system architecture.

We find that we can split our systems into two distinct evolutionary pathways.
\textit{Quiet evolution} systems have low initial multiplicities (median 4) and wide orbital spacing (median 8.2 mutual Hill radii).
These systems remain stable, ejecting a median of $28.4\%$ of their planetesimals with a median velocity of $1.611$ km s$^{-1}$.
In contrast, the \textit{catastrophic evolution} systems have high initial multiplicities (median 6), and more compact orbital architectures (median 5.9 mutual Hill radii).
The systems in this cluster eject a median $59\%$ of their planetesimals, at higher velocities ($2.9$ km s$^{-1}$).
The two populations undergo different evolutionary pathways, with the catastrophic systems ejecting a median 89.4 M$_\oplus$ of planetary mass, and increasing orbital separations by eight mutual Hill radii.
The {\em quiet} systems have minimal orbital changes, retaining their planetary architectures.

Within each population, we further divided the clusters to understand the mechanisms separating the populations.
Catastrophic systems split into three classes: \textit{monarchic ejectors}, where a single massive planet dominates scattering, producing large ejection fractions (66\%) at low velocities; \textit{cascading ejectors}, where similarly-sized planets in tightly-packed configurations trigger violent instabilities, with many orbital crossings and high energy ejections; and finally, an \textit{intermediate ejector} subcluster which falls in the middle group between these extremes.
The quiet systems were split into \textit{stable systems}, which undergo minimal orbital evolution, and \textit{gentle ejectors}, which produce moderate ISO populations through orbital diffusion, similar to the monarchic ejectors but without enough mass for efficient scattering.
These classifications demonstrate that initial system architectures --- most significantly mutiplicity, mass partitioning, total mass, and orbital spacing --- determine the efficiency and kinematics of a system's ISO ejections.

We compared our results with a longer-running Solar System simulation from \citetalias{nesvorny2023} to show that our results capture the early, instability-driven phase of ISO production.
Our velocity distributions at 10 Myr were similar to the Solar System simulation at the same epoch. 
The Solar System ISO velocity distribution develops a low-velocity tail between 1-100 Myrs, driven by the Galactic tide and stellar flybys.

Our results have several implications for the Galactic ISO population.
The relatively slow ejection velocities we find here indicate that ISOs will initially orbit in the Galaxy in dense, narrow tidal streams with small internal velocity distributions, until they are heated by inhomogeneities in the Galactic potential. 
All else equal, this fact increases the likelihood of detecting multiple ISOs from the same progenitor system. 
At the same time, the comparatively low ejection velocities we find here provides more evidence for the common assumption that the velocity distribution function of ISOs that the Solar System encounters will be set mostly by the distribution function of the stars.
%The velocity distributions we find provide a potential diagnostic of source architectures, especially when combined with compositional information which may provide information on the ISO age.

While our simulations omit several effects --- Galactic tides, collisions and planetary radii effects, and self-stirring of the planetesimal disc --- they isolate the fundamental gravitational dynamics between the planets, and demonstrate how system architectures impact ISO production efficiency and velocity distributions.
The scale-free nature of gravity also allows our results to be rescaled to systems of other masses or orbital radii.
Future work incorporating longer integration times, more physically-motivated planetary architectures, and simulations placed in their Galactic context, will help to refine predictions and assess how robust they are across a broader swath of the planetary system parameter space.

If dynamical instabilities are a common process, this would be the main release mechanism for the Galactic ISO population.
The properties of ejected ISOs retain a measurable imprint of their parent system’s architecture and age --- creating the possibility, through challenging, of using ISOs as a form of Galactic archaeology.
As our observed sample of ISOs grows (e.g. through the Vera C. Rubin Observatory's LSST, predicted to find $\sim$1 ISO per year by \citet{Dorsey2025}), the detected population will provide tests to the mechanisms and galactic propagation we predict.
Understanding this link is key to interpreting the growing observational catalogue of interstellar interlopers and reconstructing the dynamical histories of planetary systems throughout the Galaxy.

\section*{Acknowledgments}

M.T.B. and J.C.F. appreciate support by the Rutherford Discovery Fellowships from New Zealand Government funding, administered by the Royal Society Te Ap\={a}rangi. 
We are grateful to Chris Lintott for thoughtful feedback on the manuscript, and to Susanne Pfalzner, Simon Portegies Zwart, Matthew Hopkins, W. Garrett Levine, Joe Masiero, and Matthew Holman for helpful conversations.

%% To help institutions obtain information on the effectiveness of their 
%% telescopes the AAS Journals has created a group of keywords for telescope 
%% facilities.
%
%% Following the acknowledgments section, use the following syntax and the
%% \facility{} or \facilities{} macros to list the keywords of facilities used 
%% in the research for the paper.  Each keyword is check against the master 
%% list during copy editing.  Individual instruments can be provided in 
%% parentheses, after the keyword, but they are not verified.

% \vspace{5mm}
% \facilities{HST(STIS), Swift(XRT and UVOT), AAVSO, CTIO:1.3m,
% CTIO:1.5m,CXO}

%% Similar to \facility{}, there is the optional \software command to allow 
%% authors a place to specify which programs were used during the creation of 
%% the manuscript. Authors should list each code and include either a
%% citation or url to the code inside ()s when available.

Simulations were carried out using \code{rebound} \citep{rebound, reinwh2015, rein2019, simulationarchive}. Data analysis was performed using \code{NumPy} \citep{numpy}, \code{pandas} \citep{pandas1, pandas2} \code{Astropy} \citep{astropy:2013, astropy:2018, astropy:2022}, \code{scikit-learn} \citep{scikit-learn}, \code{dynesty} \citep{sergey_koposov_2024_12537467}, \code{SciPy} \citep{2020SciPy-NMeth}. Graphs were produced using \code{Matplotlib} \citep{Hunter2007}.

%% Appendix material should be preceded with a single \appendix command.
%% There should be a \section command for each appendix. Mark appendix
%% subsections with the same markup you use in the main body of the paper.

%% Each Appendix (indicated with \section) will be lettered A, B, C, etc.
%% The equation counter will reset when it encounters the \appendix
%% command and will number appendix equations (A1), (A2), etc. The
%% Figure and Table counter will not reset.

\appendix

\begin{figure}
    \centering
    \includegraphics{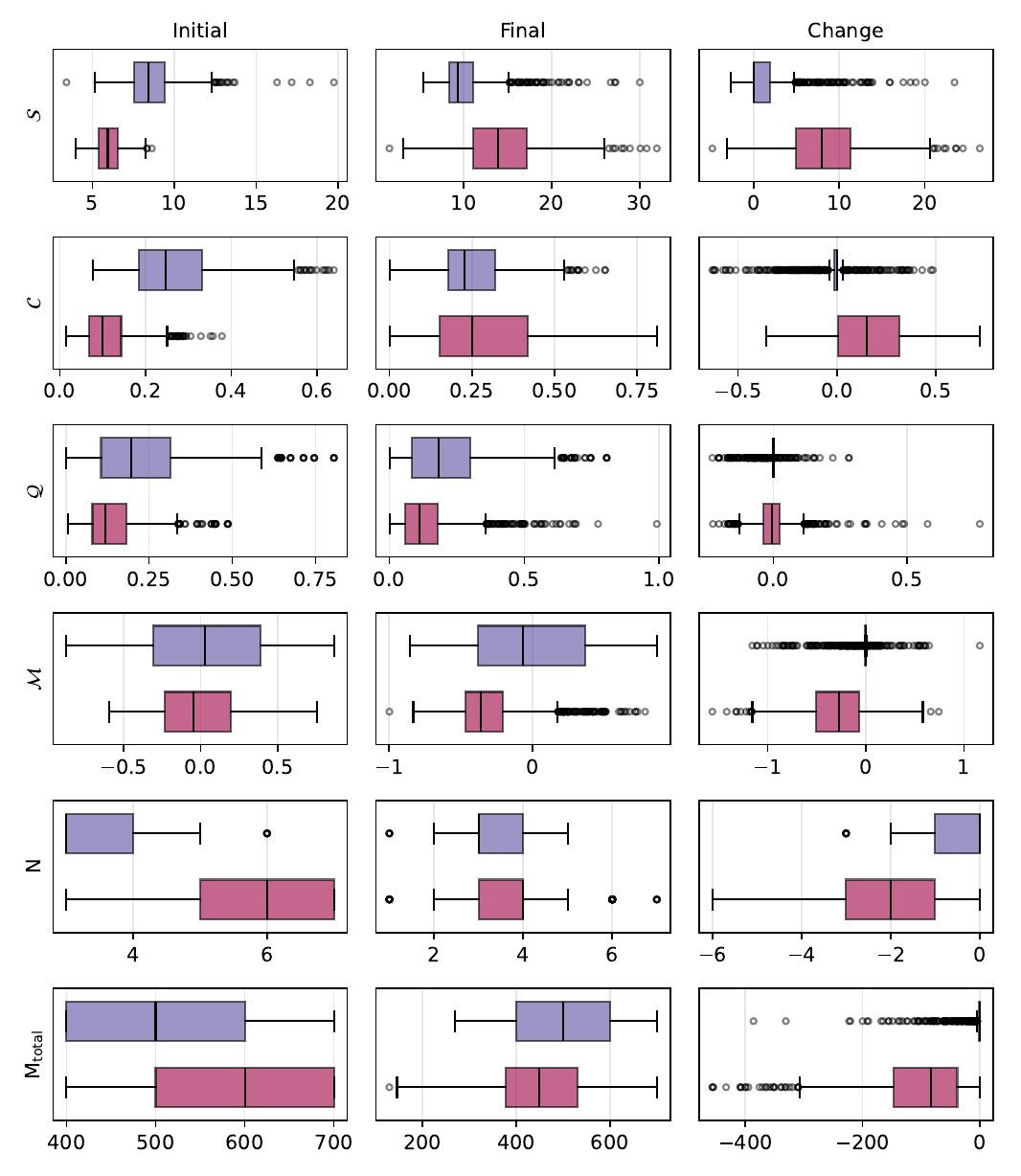}
    \caption{Box plots showing the spread of the system architecture parameters for the each broad cluster. 
    Parameter distributions for \textit{catastrophic evolution} systems are shown in red, and the \textit{quiet evolution} systems are shown in blue.}
    \label{fig:cluster_hists2}
\end{figure}

\begin{figure}
    \centering
    \includegraphics{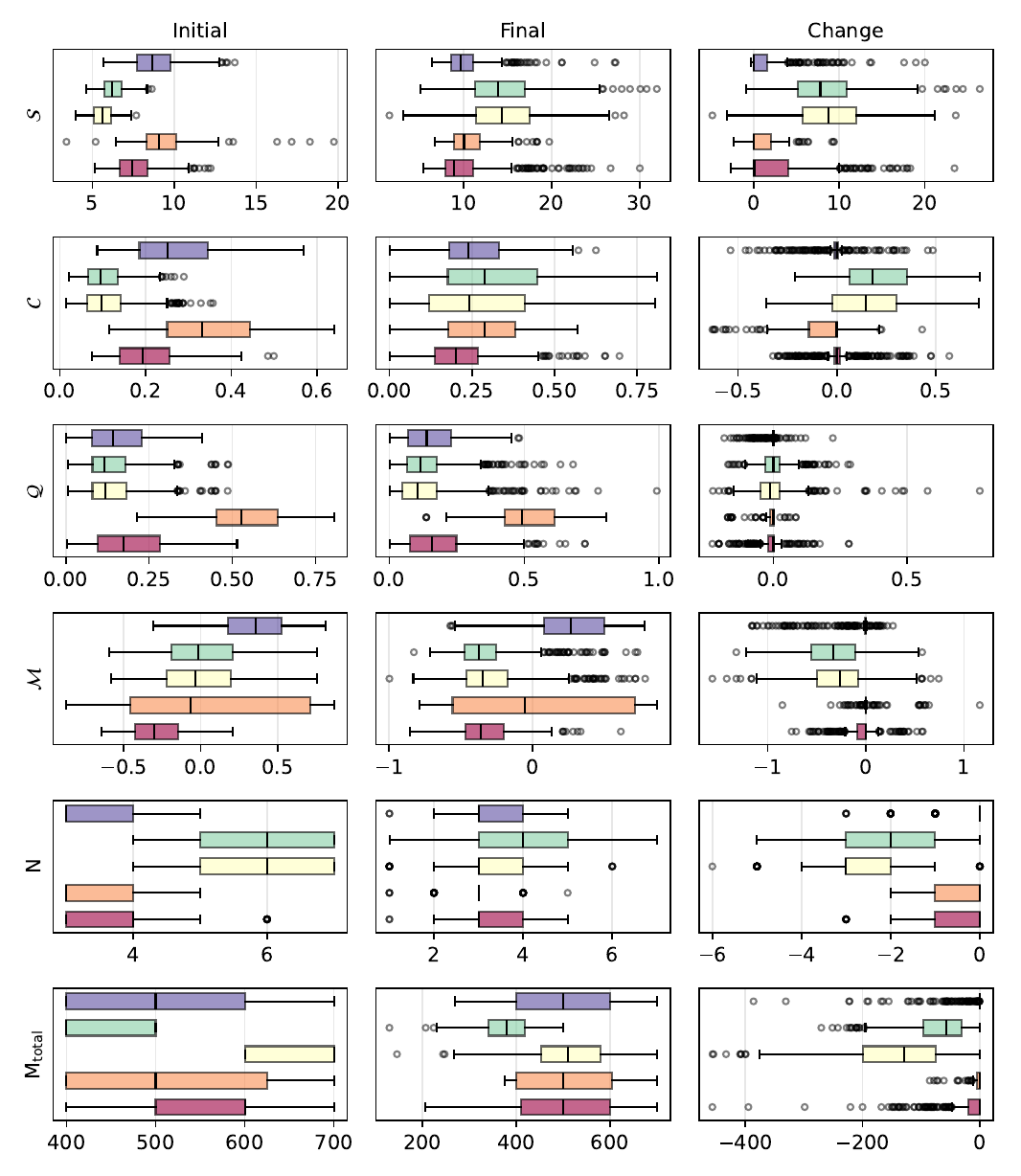}
    \caption{Box plots showing the spread of system architecture parameters for the subclusters.
    The plots match the colours in \autoref{fig:clusters}, with gentle (red) and stable (orange) the subclusters for the \textit{quiet evolution} cluster, and monarchic (blue), intermediate (green) and cascading (yellow) the subclusters for the \textit{catastrophic evolution} cluster.}
    \label{fig:cluster_hists5}
\end{figure}

The box and whisker plots for the data presented in \autoref{tab:sim_results} and \autoref{tab:subclusters} are presented in \autoref{fig:cluster_hists2} and \autoref{fig:cluster_hists5} respectively. These are the distributions of the system architecture parameters within each of the clusters found in the $k=2$ and $k=5$ $k-$means clustering analyses.
 
\bibliography{references}{}

@ARTICLE{Bhaskar:2025,
       author = {{Gautham Bhaskar}, Hareesh and {Perets}, Hagai},
        title = "{Properties of Free Floating Planets Ejected through Planet-Planet Scattering}",
      journal = {arXiv e-prints},
         year = 2025,
        month = jan,
          eid = {arXiv:2501.13166},
        pages = {arXiv:2501.13166},
          doi = {10.48550/arXiv.2501.13166},
archivePrefix = {arXiv},
       eprint = {2501.13166},
 primaryClass = {astro-ph.EP},
       adsurl = {https://ui.adsabs.harvard.edu/abs/2025arXiv250113166G}
}

@ARTICLE{Forbes2024,
       author = {{Forbes}, John C. and {Bannister}, Michele T. and {Lintott}, Chris and {Forrest}, Angus and {Portegies Zwart}, Simon and {Dorsey}, Rosemary C. and {Albrow}, Leah and {Hopkins}, Matthew J.},
        title = "{He Awa Whiria: The Tidal Streams of Interstellar Objects}",
      journal = {\apj},
         year = 2025,
        month = jul,
       volume = {988},
       number = {1},
          eid = {121},
        pages = {121},
          doi = {10.3847/1538-4357/adc9ac},
archivePrefix = {arXiv},
       eprint = {2411.14577},
 primaryClass = {astro-ph.EP},
       adsurl = {https://ui.adsabs.harvard.edu/abs/2025ApJ...988..121F}
}

@ARTICLE{Cordiner2025,
       author = {{Cordiner}, Martin A. and {Roth}, Nathan X. and {Kelley}, Michael S.~P. and {Bodewits}, Dennis and {Charnley}, Steven B. and {Drozdovskaya}, Maria N. and {Farnocchia}, Davide and {Micheli}, Marco and {Milam}, Stefanie N. and {Opitom}, Cyrielle and {Schwamb}, Megan E. and {Thomas}, Cristina A. and {Bagnulo}, Stefano},
        title = "{JWST Detection of a Carbon-dioxide-dominated Gas Coma Surrounding Interstellar Object 3I/ATLAS}",
      journal = {\apjl},
         year = 2025,
        month = oct,
       volume = {991},
       number = {2},
          eid = {L43},
        pages = {L43},
          doi = {10.3847/2041-8213/ae0647},
archivePrefix = {arXiv},
       eprint = {2508.18209},
 primaryClass = {astro-ph.EP},
       adsurl = {https://ui.adsabs.harvard.edu/abs/2025ApJ...991L..43C}
}

@ARTICLE{Holmberg2009,
       author = {{Holmberg}, J. and {Nordstr{\"o}m}, B. and {Andersen}, J.},
        title = "{The Geneva-Copenhagen survey of the solar neighbourhood. III. Improved distances, ages, and kinematics}",
      journal = {\aap},
         year = 2009,
        month = jul,
       volume = {501},
       number = {3},
        pages = {941-947},
          doi = {10.1051/0004-6361/200811191},
archivePrefix = {arXiv},
       eprint = {0811.3982},
 primaryClass = {astro-ph},
       adsurl = {https://ui.adsabs.harvard.edu/abs/2009A&A...501..941H}
}

@ARTICLE{Dorsey2025,
       author = {{Dorsey}, Rosemary C. and {Hopkins}, Matthew J. and {Bannister}, Michele T. and {Lawler}, Samantha M. and {Lintott}, Chris and {Parker}, Alex H. and {Forbes}, John C.},
        title = "{The Visibility of the {\={O}}tautahi{\textendash}Oxford Interstellar Object Population Model in LSST}",
      journal = {Planetary Science Journal},
         year = 2025,
        month = sep,
       volume = {6},
       number = {9},
          eid = {214},
        pages = {214},
          doi = {10.3847/PSJ/adf8ca},
archivePrefix = {arXiv},
       eprint = {2502.16741},
 primaryClass = {astro-ph.EP},
       adsurl = {https://ui.adsabs.harvard.edu/abs/2025PSJ.....6..214D}
}

@ARTICLE{Zhu2021,
       author = {{Zhu}, Wei and {Dong}, Subo},
        title = "{Exoplanet Statistics and Theoretical Implications}",
      journal = {\araa},
         year = 2021,
        month = sep,
       volume = {59},
        pages = {291-336},
          doi = {10.1146/annurev-astro-112420-020055},
archivePrefix = {arXiv},
       eprint = {2103.02127},
 primaryClass = {astro-ph.EP},
       adsurl = {https://ui.adsabs.harvard.edu/abs/2021ARA&A..59..291Z}
}

@article{lintott2021,
  title = {Predicting the {{Water Content}} of {{Interstellar Objects}} from {{Galactic Star Formation Histories}}},
  author = {Lintott, Chris and Bannister, Michele T. and Mackereth, J. Ted},
  year = 2021,
  month = dec,
  journal = {The Astrophysical Journal Letters},
  volume = {924},
  number = {1},
  pages = {L1},
  publisher = {The American Astronomical Society},
  issn = {2041-8205},
  doi = {10.3847/2041-8213/ac41d5},
  urldate = {2025-11-29},
  langid = {english}
}

@article{stern1990,
  title = {On the Number Density of Interstellar Comets as a Constraint on the Formation Rate of Planetary Systems},
  author = {Stern, S. Alan},
  year = 1990,
  month = jul,
  journal = {Publications of the Astronomical Society of the Pacific},
  volume = {102},
  pages = {793},
  issn = {0004-6280, 1538-3873},
  doi = {10.1086/132704},
  urldate = {2025-11-29},
  langid = {english}
}

@ARTICLE{spitzer1951,
       author = {{Spitzer}, Jr., Lyman and {Schwarzschild}, Martin},
        title = "{The Possible Influence of Interstellar Clouds on Stellar Velocities.}",
      journal = {\apj},
         year = 1951,
        month = nov,
       volume = {114},
        pages = {385},
          doi = {10.1086/145478},
       adsurl = {https://ui.adsabs.harvard.edu/abs/1951ApJ...114..385S}
}

@ARTICLE{lacey1984,
       author = {{Lacey}, C.~G.},
        title = "{The influence of massive gas clouds on stellar velocity dispersions in galactic discs}",
      journal = {\mnras},
         year = 1984,
        month = jun,
       volume = {208},
        pages = {687-707},
          doi = {10.1093/mnras/208.4.687},
       adsurl = {https://ui.adsabs.harvard.edu/abs/1984MNRAS.208..687L}
}

@ARTICLE{Forbes2012,
       author = {{Forbes}, John and {Krumholz}, Mark and {Burkert}, Andreas},
        title = "{Evolving Gravitationally Unstable Disks over Cosmic Time: Implications for Thick Disk Formation}",
      journal = {\apj},
         year = 2012,
        month = jul,
       volume = {754},
       number = {1},
          eid = {48},
        pages = {48},
          doi = {10.1088/0004-637X/754/1/48},
archivePrefix = {arXiv},
       eprint = {1112.1410},
 primaryClass = {astro-ph.GA},
       adsurl = {https://ui.adsabs.harvard.edu/abs/2012ApJ...754...48F}
}

@ARTICLE{Huang2025,
       author = {{Huang}, Yukun and {Gladman}, Brett and {Kokubo}, Eiichiro},
        title = "{Analytical Solutions for Planet-Scattering Small Bodies}",
      journal = {arXiv e-prints},
         year = 2025,
        month = nov,
          eid = {arXiv:2511.16056},
        pages = {arXiv:2511.16056},
          doi = {10.48550/arXiv.2511.16056},
archivePrefix = {arXiv},
       eprint = {2511.16056},
 primaryClass = {astro-ph.EP},
       adsurl = {https://ui.adsabs.harvard.edu/abs/2025arXiv251116056H}
}

@ARTICLE{Hadden2024,
       author = {{Hadden}, Sam and {Tremaine}, Scott},
        title = "{Scattered disc dynamics: the mapping approach}",
      journal = {\mnras},
         year = 2024,
        month = jan,
       volume = {527},
       number = {2},
        pages = {3054-3075},
          doi = {10.1093/mnras/stad3478},
archivePrefix = {arXiv},
       eprint = {2309.00684},
 primaryClass = {astro-ph.EP},
       adsurl = {https://ui.adsabs.harvard.edu/abs/2024MNRAS.527.3054H}
}

@ARTICLE{Carlberg2024,
       author = {{Carlberg}, Raymond G. and {Jenkins}, Adrian and {Frenk}, Carlos S. and {Cooper}, Andrew P.},
        title = "{Star Stream Velocity Distributions in Cold Dark Matter and Warm Dark Matter Galactic Halos}",
      journal = {\apj},
         year = 2024,
        month = nov,
       volume = {975},
       number = {1},
          eid = {135},
        pages = {135},
          doi = {10.3847/1538-4357/ad7b35},
archivePrefix = {arXiv},
       eprint = {2405.18522},
 primaryClass = {astro-ph.GA},
       adsurl = {https://ui.adsabs.harvard.edu/abs/2024ApJ...975..135C}
}

@ARTICLE{Hopkins2025,
       author = {{Hopkins}, Matthew J. and {Bannister}, Michele T. and {Lintott}, Chris},
        title = "{Predicting Interstellar Object Chemodynamics with Gaia}",
      journal = {\aj},
         year = 2025,
        month = feb,
       volume = {169},
       number = {2},
          eid = {78},
        pages = {78},
          doi = {10.3847/1538-3881/ad9eb3},
archivePrefix = {arXiv},
       eprint = {2402.04904},
 primaryClass = {astro-ph.EP},
       adsurl = {https://ui.adsabs.harvard.edu/abs/2025AJ....169...78H}
}

@ARTICLE{Gaia2018,
       author = {{Gaia Collaboration} and {Katz}, D. and {Antoja}, T. and {Romero-G{\'o}mez}, M. and {Drimmel}, R. and {Reyl{\'e}}, C. and {Seabroke}, G.~M. and {Soubiran}, C. and {Babusiaux}, C. and {Di Matteo}, P. and {Figueras}, F. and {Poggio}, E. and {Robin}, A.~C. and {Evans}, D.~W. and {Brown}, A.~G.~A. and {Vallenari}, A. and {Prusti}, T. and {de Bruijne}, J.~H.~J. and {Bailer-Jones}, C.~A.~L. and {Biermann}, M. and {Eyer}, L. and {Jansen}, F. and {Jordi}, C. and {Klioner}, S.~A. and {Lammers}, U. and {Lindegren}, L. and {Luri}, X. and {Mignard}, F. and {Panem}, C. and {Pourbaix}, D. and {Randich}, S. and {Sartoretti}, P. and {Siddiqui}, H.~I. and {van Leeuwen}, F. and {Walton}, N.~A. and {Arenou}, F. and {Bastian}, U. and {Cropper}, M. and {Lattanzi}, M.~G. and {Bakker}, J. and {Cacciari}, C. and {Casta n}, J. and {Chaoul}, L. and {Cheek}, N. and {De Angeli}, F. and {Fabricius}, C. and {Guerra}, R. and {Holl}, B. and {Masana}, E. and {Messineo}, R. and {Mowlavi}, N. and {Nienartowicz}, K. and {Panuzzo}, P. and {Portell}, J. and {Riello}, M. and {Tanga}, P. and {Th{\'e}venin}, F. and {Gracia-Abril}, G. and {Comoretto}, G. and {Garcia-Reinaldos}, M. and {Teyssier}, D. and {Altmann}, M. and {Andrae}, R. and {Audard}, M. and {Bellas-Velidis}, I. and {Benson}, K. and {Berthier}, J. and {Blomme}, R. and {Burgess}, P. and {Busso}, G. and {Carry}, B. and {Cellino}, A. and {Clementini}, G. and {Clotet}, M. and {Creevey}, O. and {Davidson}, M. and {De Ridder}, J. and {Delchambre}, L. and {Dell'Oro}, A. and {Ducourant}, C. and {Fern{\'a}ndez-Hern{\'a}ndez}, J. and {Fouesneau}, M. and {Fr{\'e}mat}, Y. and {Galluccio}, L. and {Garc{\'\i}a-Torres}, M. and {Gonz{\'a}lez-N{\'u}{\~n}ez}, J. and {Gonz{\'a}lez-Vidal}, J.~J. and {Gosset}, E. and {Guy}, L.~P. and {Halbwachs}, J. -L. and {Hambly}, N.~C. and {Harrison}, D.~L. and {Hern{\'a}ndez}, J. and {Hestroffer}, D. and {Hodgkin}, S.~T. and {Hutton}, A. and {Jasniewicz}, G. and {Jean-Antoine-Piccolo}, A. and {Jordan}, S. and {Korn}, A.~J. and {Krone-Martins}, A. and {Lanzafame}, A.~C. and {Lebzelter}, T. and {L{\"o}ffler}, W. and {Manteiga}, M. and {Marrese}, P.~M. and {Mart{\'\i}n-Fleitas}, J.~M. and {Moitinho}, A. and {Mora}, A. and {Muinonen}, K. and {Osinde}, J. and {Pancino}, E. and {Pauwels}, T. and {Petit}, J. -M. and {Recio-Blanco}, A. and {Richards}, P.~J. and {Rimoldini}, L. and {Sarro}, L.~M. and {Siopis}, C. and {Smith}, M. and {Sozzetti}, A. and {S{\"u}veges}, M. and {Torra}, J. and {van Reeven}, W. and {Abbas}, U. and {Abreu Aramburu}, A. and {Accart}, S. and {Aerts}, C. and {Altavilla}, G. and {{\'A}lvarez}, M.~A. and {Alvarez}, R. and {Alves}, J. and {Anderson}, R.~I. and {Andrei}, A.~H. and {Anglada Varela}, E. and {Antiche}, E. and {Arcay}, B. and {Astraatmadja}, T.~L. and {Bach}, N. and {Baker}, S.~G. and {Balaguer-N{\'u}{\~n}ez}, L. and {Balm}, P. and {Barache}, C. and {Barata}, C. and {Barbato}, D. and {Barblan}, F. and {Barklem}, P.~S. and {Barrado}, D. and {Barros}, M. and {Barstow}, M.~A. and {Bartholom{\'e} Mu{\~n}oz}, L. and {Bassilana}, J. -L. and {Becciani}, U. and {Bellazzini}, M. and {Berihuete}, A. and {Bertone}, S. and {Bianchi}, L. and {Bienaym{\'e}}, O. and {Blanco-Cuaresma}, S. and {Boch}, T. and {Boeche}, C. and {Bombrun}, A. and {Borrachero}, R. and {Bossini}, D. and {Bouquillon}, S. and {Bourda}, G. and {Bragaglia}, A. and {Bramante}, L. and {Breddels}, M.~A. and {Bressan}, A. and {Brouillet}, N. and {Br{\"u}semeister}, T. and {Brugaletta}, E. and {Bucciarelli}, B. and {Burlacu}, A. and {Busonero}, D. and {Butkevich}, A.~G. and {Buzzi}, R. and {Caffau}, E. and {Cancelliere}, R. and {Cannizzaro}, G. and {Cantat-Gaudin}, T. and {Carballo}, R. and {Carlucci}, T. and {Carrasco}, J.~M. and {Casamiquela}, L. and {Castellani}, M. and {Castro-Ginard}, A. and {Charlot}, P. and {Chemin}, L. and {Chiavassa}, A. and {Cocozza}, G. and {Costigan}, G. and {Cowell}, S. and {Crifo}, F. and {Crosta}, M. and {Crowley}, C. and {Cuypers}, J. and {Dafonte}, C. and {Damerdji}, Y.},
        title = "{Gaia Data Release 2. Mapping the Milky Way disc kinematics}",
      journal = {\aap},
         year = 2018,
        month = aug,
       volume = {616},
          eid = {A11},
        pages = {A11},
          doi = {10.1051/0004-6361/201832865},
archivePrefix = {arXiv},
       eprint = {1804.09380},
 primaryClass = {astro-ph.GA},
       adsurl = {https://ui.adsabs.harvard.edu/abs/2018A&A...616A..11G}
}

@ARTICLE{Opitom2025,
       author = {{Opitom}, Cyrielle and {Snodgrass}, Colin and {Jehin}, Emmanuel and {Bannister}, Michele T. and {Bufanda}, Erica and {Deam}, Sophie E. and {Dorsey}, Rosemary C. and {Ferrais}, Marin and {Hmiddouch}, Said and {Knight}, Matthew M. and {Kokotanekova}, Rosita and {Leicester}, Brayden and {Marsset}, Micha{\"e}l and {Murphy}, Brian and {Okoth}, Vincent and {Ridden-Harper}, Ryan and {Vander Donckt}, Mathieu and {Ferellec}, L{\'e}a and {Hutsem{\'e}kers}, Damien and {Lippi}, Manuela and {Manfroid}, Jean and {Benkhaldoun}, Zouhair},
        title = "{Snapshot of a new interstellar comet: 3I/ATLAS has a red and featureless spectrum}",
      journal = {\mnras},
         year = 2025,
        month = nov,
       volume = {544},
       number = {1},
        pages = {L31-L36},
          doi = {10.1093/mnrasl/slaf095},
archivePrefix = {arXiv},
       eprint = {2507.05226},
 primaryClass = {astro-ph.EP},
       adsurl = {https://ui.adsabs.harvard.edu/abs/2025MNRAS.544L..31O}
}

@article{Andrews2020,
  title = {Observations of {{Protoplanetary Disk Structures}}},
  author = {Andrews, Sean M.},
  year = {2020},
  month = aug,
  journal = {Annual Review of Astronomy and Astrophysics},
  volume = {58},
  number = {Volume 58, 2020},
  eprint = {2001.05007},
  pages = {483--528},
  publisher = {Annual Reviews Inc.},
  issn = {00664146},
  doi = {10.1146/annurev-astro-031220-010302},
  urldate = {2024-04-27},
  archiveprefix = {arXiv}
}

@incollection{Levison2007,
  title = {Planet {{Migration}} in {{Planetesimal Disks}}},
  booktitle = {Protostars and {{Planets V}}},
  author = {Levison, Harold F and Morbidelli, Alessandro and Nacional, Observat{\'o}rio and De Janeiro, Rio and Backman, Dana},
  year = {2007},
  pages = {669--684},
  publisher = {University of Arizona Press}
}

@ARTICLE{Dehnen2018,
       author = {{Dehnen}, Walter and {Hasanuddin}},
        title = "{Tidal ribbons}",
      journal = {\mnras},
         year = 2018,
        month = oct,
       volume = {479},
       number = {4},
        pages = {4720-4726},
          doi = {10.1093/mnras/sty1726},
archivePrefix = {arXiv},
       eprint = {1805.08481},
 primaryClass = {astro-ph.GA},
       adsurl = {https://ui.adsabs.harvard.edu/abs/2018MNRAS.479.4720D}
}

@ARTICLE{Taylor2025,
       author = {{Taylor}, Aster G. and {Seligman}, Darryl Z.},
        title = "{The Kinematic Age of 3I/ATLAS and Its Implications for Early Planet Formation}",
      journal = {\apjl},
         year = 2025,
        month = sep,
       volume = {990},
       number = {1},
          eid = {L14},
        pages = {L14},
          doi = {10.3847/2041-8213/adfa28},
archivePrefix = {arXiv},
       eprint = {2507.08111},
 primaryClass = {astro-ph.EP},
       adsurl = {https://ui.adsabs.harvard.edu/abs/2025ApJ...990L..14T}
}

@ARTICLE{Hopkins2025b,
       author = {{Hopkins}, Matthew J. and {Dorsey}, Rosemary C. and {Forbes}, John C. and {Bannister}, Michele T. and {Lintott}, Chris J. and {Leicester}, Brayden},
        title = "{From a Different Star: 3I/ATLAS in the Context of the {\={O}}tautahi{\textendash}Oxford Interstellar Object Population Model}",
      journal = {\apjl},
         year = 2025,
        month = sep,
       volume = {990},
       number = {2},
          eid = {L30},
        pages = {L30},
          doi = {10.3847/2041-8213/adfbf4},
archivePrefix = {arXiv},
       eprint = {2507.05318},
 primaryClass = {astro-ph.EP},
       adsurl = {https://ui.adsabs.harvard.edu/abs/2025ApJ...990L..30H}
}

@ARTICLE{Perez-Couto2025,
       author = {{P{\'e}rez-Couto}, X. and {Torres}, S. and {Villaver}, E. and {Mustill}, A.~J. and {Manteiga}, M.},
        title = "{3I/ATLAS: In Search of the Witnesses to Its Voyage}",
      journal = {arXiv e-prints},
         year = 2025,
        month = sep,
          eid = {arXiv:2509.07678},
        pages = {arXiv:2509.07678},
          doi = {10.48550/arXiv.2509.07678},
archivePrefix = {arXiv},
       eprint = {2509.07678},
 primaryClass = {astro-ph.EP},
       adsurl = {https://ui.adsabs.harvard.edu/abs/2025arXiv250907678P}
}

@ARTICLE{Deam2025,
       author = {{Deam}, Sophie E. and {Bannister}, Michele T. and {Opitom}, Cyrielle and {Knight}, Matthew M. and {Ridden-Harper}, Ryan and {Seligman}, Darryl Z. and {Fitzsimmons}, Alan and {Guilbert-Lepoutre}, Aur{\'e}lie and {Jehin}, Emmanuel and {Jorda}, Laurent and {Marsset}, Michael and {Moulane}, Youssef and {Rousselot}, Philippe and {Vernazza}, Pierre and {Yang}, Bin},
        title = "{A portrait throughout perihelion of the NH$_2$-rich interstellar comet 2I/Borisov}",
      journal = {arXiv e-prints},
         year = 2025,
        month = jul,
          eid = {arXiv:2507.05051},
        pages = {arXiv:2507.05051},
          doi = {10.48550/arXiv.2507.05051},
archivePrefix = {arXiv},
       eprint = {2507.05051},
 primaryClass = {astro-ph.EP},
       adsurl = {https://ui.adsabs.harvard.edu/abs/2025arXiv250705051D}
}

@article{Pfalzner2021,
  title = {Significant Interstellar Object Production by Close Stellar Flybys},
  author = {Pfalzner, Susanne and Aizpuru Vargas, Luis L. and Bhandare, Asmita and Veras, {\relax Di}mitri},
  year = {2021},
  month = jul,
  journal = {Astronomy \& Astrophysics},
  volume = {651},
  eprint = {2104.06845},
  pages = {A38},
  publisher = {EDP Sciences},
  issn = {0004-6361},
  doi = {10.1051/0004-6361/202140587},
  url = {https://www.aanda.org/articles/aa/full_html/2021/07/aa40587-21/aa40587-21.html},
  urldate = {2024-04-27},
  archiveprefix = {arxiv}
}

@article{liuEarlySolarSystem2022,
  title = {Early {{Solar System}} Instability Triggered by Dispersal of the Gaseous Disk},
  author = {Liu, Beibei and Raymond, Sean N. and Jacobson, Seth A.},
  year = {2022},
  month = apr,
  journal = {Nature},
  volume = {604},
  number = {7907},
  eprint = {2205.02026},
  pages = {643--646},
  publisher = {Nature Publishing Group},
  issn = {1476-4687},
  doi = {10.1038/s41586-022-04535-1},
  url = {https://www.nature.com/articles/s41586-022-04535-1},
  urldate = {2024-04-11},
  archiveprefix = {arxiv},
  pmid = {35478235}
}

@inproceedings{safronov1972,
  title = {Ejection of {{Bodies}} from the {{Solar System}} in the {{Course}} of the {{Accumulation}} of the {{Giant Planets}} and the {{Formation}} of the {{Cometary Cloud}}},
  booktitle = {The {{Motion}}, {{Evolution}} of {{Orbits}}, and {{Origin}} of {{Comets}}},
  author = {Safronov, V. S.},
  year = {1972},
  month = jan,
  volume = {45},
  address = {Leningrad, U.S.S.R.}
}

@article{rebound,
  title = {{{REBOUND}}: An Open-Source Multi-Purpose {{N-body}} Code for Collisional Dynamics},
  shorttitle = {{{REBOUND}}},
  author = {Rein, H. and Liu, S.-F.},
  year = {2012},
  month = jan,
  journal = {Astronomy \& Astrophysics},
  volume = {537},
  pages = {A128},
  publisher = {EDP Sciences},
  issn = {0004-6361, 1432-0746},
  doi = {10.1051/0004-6361/201118085},
  url = {https://www.aanda.org/articles/aa/abs/2012/01/aa18085-11/aa18085-11.html},
  urldate = {2024-04-30},
  copyright = {{\copyright} ESO, 2012},
  langid = {english}
}

@article{wisdom1991,
  title = {Symplectic Maps for the {{N-body}} Problem},
  author = {Wisdom, Jack and Holman, Matthew},
  year = {1991},
  month = oct,
  journal = {The Astronomical Journal},
  volume = {102},
  pages = {1528--1538},
  doi = {10.1086/115978},
  url = {https://ui.adsabs.harvard.edu/abs/1991AJ....102.1528W},
  langid = {english}
}

@article{Raymond2020,
  title = {Survivor {{Bias}}: {{Divergent Fates}} of the {{Solar System}}'s {{Ejected}} versus {{Persisting Planetesimals}}},
  author = {Raymond, Sean N and Kaib, Nathan A and Armitage, Philip J and Fortney, Jonathan J},
  year = {2020},
  journal = {The Astrophysical Journal Letters},
  doi = {10.3847/2041-8213/abc55f},
  url = {https://doi.org/10.3847/2041-8213/abc55f},
  urldate = {2024-04-04}
}

@article{reinwh2015,
  title = {{{WHFAST}}: {{A}} Fast and Unbiased Implementation of a Symplectic {{Wisdom-Holman}} Integrator for Long-Term Gravitational Simulations},
  author = {Rein, Hanno and Tamayo, Daniel},
  year = {2015},
  month = apr,
  journal = {Monthly Notices of the Royal Astronomical Society},
  volume = {452},
  number = {1},
  eprint = {1506.01084},
  pages = {376--388},
  publisher = {Oxford University Press},
  issn = {13652966},
  doi = {10.1093/mnras/stv1257},
  archiveprefix = {arxiv}
}

@article{gilbert2020,
  title = {An {{Information Theoretic Framework}} for {{Classifying Exoplanetary System Architectures}}},
  author = {Gilbert, Gregory J. and Fabrycky, Daniel C.},
  year = {2020},
  month = may,
  journal = {The Astronomical Journal},
  volume = {159},
  number = {6},
  eprint = {2003.11098},
  pages = {281},
  publisher = {IOP Publishing},
  issn = {1538-3881},
  doi = {10.3847/1538-3881/AB8E3C},
  url = {https://iopscience.iop.org/article/10.3847/1538-3881/ab8e3c},
  urldate = {2024-04-11},
  archiveprefix = {arxiv}
}

@article{laughlin2017,
  title = {On the {{Consequences}} of the {{Detection}} of an {{Interstellar Asteroid}}},
  author = {Laughlin, Gregory and Batygin, Konstantin},
  year = {2017},
  month = dec,
  journal = {Research Notes of the AAS},
  volume = {1},
  number = {1},
  pages = {43},
  issn = {2515-5172},
  doi = {10.3847/2515-5172/aaa02b},
  url = {https://dx.doi.org/10.3847/2515-5172/aaa02b},
  langid = {english}
}

@article{levine2023,
  title = {Interstellar {{Comets}} from {{Post-main-sequence Systems}} as {{Tracers}} of {{Extrasolar Oort Clouds}}},
  author = {Levine, W. Garrett and Taylor, Aster G and Seligman, Darryl Z and Hoover, Devin J and Jedicke, Robert and Bergner, Jennifer B and Laughlin, Gregory},
  year = {2023},
  month = jul,
  journal = {Planetary Science Journal},
  volume = {4},
  number = {7},
  eprint = {2306.12464},
  publisher = {Institute of Physics},
  issn = {26323338},
  doi = {10.3847/PSJ/acdf58},
  archiveprefix = {arxiv}
}

@article{veras2014,
  title = {Hydrogen Delivery onto White Dwarfs from Remnant Exo-{{Oort}} Cloud Comets},
  author = {Veras, Dimitri and Shannon, Andrew and G{\"a}nsicke, Boris T.},
  year = {2014},
  month = dec,
  journal = {Monthly Notices of the Royal Astronomical Society},
  volume = {445},
  number = {4},
  pages = {4175--4185},
  issn = {1365-2966, 0035-8711},
  doi = {10.1093/mnras/stu2026},
  url = {http://academic.oup.com/mnras/article/445/4/4175/1752683/Hydrogen-delivery-onto-white-dwarfs-from-remnant},
  urldate = {2024-05-02},
  langid = {english}
}

@article{bannister2019,
  title = {The Natural History of `{{Oumuamua}}},
  author = {{The `Oumuamua ISSI Team} and Bannister, Michele T. and Bhandare, Asmita and Dybczy{\'n}ski, Piotr A. and Fitzsimmons, Alan and {Guilbert-Lepoutre}, Aur{\'e}lie and Jedicke, Robert and Knight, Matthew M. and Meech, Karen J. and McNeill, Andrew and Pfalzner, Susanne and Raymond, Sean N. and Snodgrass, Colin and Trilling, David E. and Ye, Quanzhi},
  year = {2019},
  month = jul,
  journal = {Nature Astronomy},
  volume = {3},
  number = {7},
  pages = {594--602},
  publisher = {Nature Publishing Group},
  issn = {2397-3366},
  doi = {10.1038/s41550-019-0816-x},
  url = {https://www.nature.com/articles/s41550-019-0816-x},
  urldate = {2024-05-29},
  copyright = {2019 Springer Nature Limited},
  langid = {english}
}

@incollection{lovis2010,
  title = {Radial {{Velocity Techniques}} for {{Exoplanets}}},
  booktitle = {Exoplanets},
  author = {Lovis, C. and Fischer, D.},
  editor = {Seager, S},
  year = {2010},
  month = dec,
  pages = {27--53},
  publisher = {University of Arizona Press},
  address = {Tucson},
  url = {https://ui.adsabs.harvard.edu/abs/2010exop.book...27L},
  urldate = {2024-05-30}
}

@article{adams2005,
  title = {Lithopanspermia in {{Star-Forming Clusters}}},
  author = {Adams, Fred C. and Spergel, David N.},
  year = {2005},
  month = aug,
  journal = {Astrobiology},
  volume = {5},
  number = {4},
  pages = {497--514},
  issn = {1531-1074, 1557-8070},
  doi = {10.1089/ast.2005.5.497},
  url = {http://www.liebertpub.com/doi/10.1089/ast.2005.5.497},
  urldate = {2024-05-02},
  langid = {english}
}

@article{eriksson2021,
  title = {The Fate of Planetesimals Formed at Planetary Gap Edges},
  author = {Eriksson, Linn E. J. and Ronnet, Thomas and Johansen, Anders},
  year = {2021},
  month = apr,
  journal = {Astronomy \& Astrophysics},
  volume = {648},
  pages = {A112},
  publisher = {EDP Sciences},
  issn = {0004-6361, 1432-0746},
  doi = {10.1051/0004-6361/202039889},
  url = {https://www.aanda.org/articles/aa/abs/2021/04/aa39889-20/aa39889-20.html},
  urldate = {2024-06-03},
  copyright = {{\copyright} ESO 2021},
  langid = {english}
}

@ARTICLE{simulationarchive,
       author = {{Rein}, Hanno and {Tamayo}, Daniel},
        title = "{A new paradigm for reproducing and analyzing N-body simulations of planetary systems}",
      journal = {\mnras},
         year = 2017,
        month = may,
       volume = {467},
       number = {2},
        pages = {2377-2383},
          doi = {10.1093/mnras/stx232},
archivePrefix = {arXiv},
       eprint = {1701.07423},
 primaryClass = {astro-ph.EP},
       adsurl = {https://ui.adsabs.harvard.edu/abs/2017MNRAS.467.2377R}
}

@ARTICLE{massradius,
       author = {{M{\"u}ller}, Simon and {Baron}, Jana and {Helled}, Ravit and {Bouchy}, Fran{\c{c}}ois and {Parc}, L{\'e}na},
        title = "{The mass-radius relation of exoplanets revisited}",
      journal = {\aap},
         year = 2024,
        month = jun,
       volume = {686},
          eid = {A296},
        pages = {A296},
          doi = {10.1051/0004-6361/202348690},
archivePrefix = {arXiv},
       eprint = {2311.12593},
 primaryClass = {astro-ph.EP},
       adsurl = {https://ui.adsabs.harvard.edu/abs/2024A&A...686A.296M}
}

@article{malmberg2011,
  title = {The Effects of Fly-Bys on Planetary Systems},
  author = {Malmberg, Daniel and Davies, Melvyn B. and Heggie, Douglas C.},
  year = {2011},
  month = feb,
  journal = {Monthly Notices of the Royal Astronomical Society},
  volume = {411},
  number = {2},
  pages = {859--877},
  issn = {0035-8711},
  doi = {10.1111/j.1365-2966.2010.17730.x},
  urldate = {2025-04-15}
}

@article{izidoro2017,
  title = {Breaking the Chains: Hot Super-{{Earth}} Systems from Migration and Disruption of Compact Resonant Chains},
  shorttitle = {Breaking the Chains},
  author = {Izidoro, Andre and Ogihara, Masahiro and Raymond, Sean N. and Morbidelli, Alessandro and Pierens, Arnaud and Bitsch, Bertram and Cossou, Christophe and Hersant, Franck},
  year = {2017},
  month = sep,
  journal = {Monthly Notices of the Royal Astronomical Society},
  volume = {470},
  number = {2},
  pages = {1750--1770},
  issn = {0035-8711},
  doi = {10.1093/mnras/stx1232},
  urldate = {2025-05-20}
}

@incollection{morbidelliDynamicalEvolutionPlanetary2018,
  title = {Dynamical {{Evolution}} of {{Planetary Systems}}},
  booktitle = {Handbook of {{Exoplanets}}},
  author = {Morbidelli, Alessandro},
  editor = {Deeg, H. and Belmonte, J.A},
  year = {2018},
  month = mar,
  eprint = {1803.06704v1},
  pages = {2523--2541},
  publisher = {Springer International Publishing},
  doi = {10.1007/978-3-319-55333-7_145},
  urldate = {2024-04-14},
  archiveprefix = {arXiv}
}

@article{moro-martin2009,
  title = {{{WILL THE LARGE SYNOPTIC SURVEY TELESCOPE DETECT EXTRA-SOLAR PLANETESIMALS ENTERING THE SOLAR SYSTEM}}?},
  author = {{Moro-Mart{\'i}n}, Amaya and Turner, Edwin L. and Loeb, Abraham},
  year = {2009},
  month = oct,
  journal = {The Astrophysical Journal},
  volume = {704},
  number = {1},
  pages = {733--742},
  issn = {0004-637X, 1538-4357},
  doi = {10.1088/0004-637X/704/1/733},
  urldate = {2025-07-30}
}

@article{bannister2018,
  title = {{{OSSOS}}. {{VII}}. 800+ {{Trans-Neptunian Objects}}---{{The Complete Data Release}}},
  author = {Bannister, Michele T. and Gladman, Brett J. and Kavelaars, J. J. and Petit, Jean-Marc and Volk, Kathryn and Chen 陳英, Ying-Tung 同 and Alexandersen, Mike and Gwyn, Stephen D. J. and Schwamb, Megan E. and Ashton, Edward and Benecchi, Susan D. and Cabral, Nahuel and Dawson, Rebekah I. and Delsanti, Audrey and Fraser, Wesley C. and Granvik, Mikael and Greenstreet, Sarah and {Guilbert-Lepoutre}, Aur{\'e}lie and Ip 葉永, Wing-Huen 烜 and Jakubik, Marian and Jones, R. Lynne and Kaib, Nathan A. and Lacerda, Pedro and Laerhoven, Christa Van and Lawler, Samantha and Lehner, Matthew J. and Lin 林省, Hsing Wen 文 and Lykawka, Patryk Sofia and Marsset, Micha{\"e}l and {Murray-Clay}, Ruth and Pike, Rosemary E. and Rousselot, Philippe and Shankman, Cory and Thirouin, Audrey and Vernazza, Pierre and Wang 王祥, Shiang-Yu 宇},
  year = {2018},
  month = may,
  journal = {The Astrophysical Journal Supplement Series},
  volume = {236},
  number = {1},
  pages = {18},
  issn = {0067-0049, 1538-4365},
  doi = {10.3847/1538-4365/aab77a},
  urldate = {2025-07-30}
}

@article{dawson2018,
  title = {Origins of {{Hot Jupiters}}},
  author = {Dawson, Rebekah I. and Johnson, John Asher},
  year = {2018},
  month = sep,
  journal = {Annual Review of Astronomy and Astrophysics},
  volume = {56},
  number = {Volume 56, 2018},
  pages = {175--221},
  publisher = {Annual Reviews},
  issn = {0066-4146, 1545-4282},
  doi = {10.1146/annurev-astro-081817-051853},
  urldate = {2025-08-07},
  langid = {english}
}

@article{rein2019,
  title = {Hybrid Symplectic Integrators for Planetary Dynamics},
  author = {Rein, Hanno and Hernandez, David M and Tamayo, Daniel and Brown, Garett and Eckels, Emily and Holmes, Emma and Lau, Michelle and Leblanc, R{\'e}jean and Silburt, Ari},
  year = {2019},
  month = jun,
  journal = {Monthly Notices of the Royal Astronomical Society},
  volume = {485},
  number = {4},
  pages = {5490--5497},
  issn = {0035-8711},
  doi = {10.1093/mnras/stz769},
  urldate = {2025-08-07}
}

@article{hopkins2023,
  title = {The {{Galactic Interstellar Object Population}}: {{A Framework}} for {{Prediction}} and {{Inference}}},
  shorttitle = {The {{Galactic Interstellar Object Population}}},
  author = {Hopkins, Matthew J. and Lintott, Chris and Bannister, Michele T. and Mackereth, J. Ted and Forbes, John C.},
  year = {2023},
  month = nov,
  journal = {The Astronomical Journal},
  volume = {166},
  number = {6},
  pages = {241},
  publisher = {The American Astronomical Society},
  issn = {1538-3881},
  doi = {10.3847/1538-3881/ad03e6},
  urldate = {2025-08-11},
  langid = {english}
}

@Article{         numpy,
 title         = {Array programming with {NumPy}},
 author        = {Charles R. Harris and K. Jarrod Millman and St{\'{e}}fan J.
                 van der Walt and Ralf Gommers and Pauli Virtanen and David
                 Cournapeau and Eric Wieser and Julian Taylor and Sebastian
                 Berg and Nathaniel J. Smith and Robert Kern and Matti Picus
                 and Stephan Hoyer and Marten H. van Kerkwijk and Matthew
                 Brett and Allan Haldane and Jaime Fern{\'{a}}ndez del
                 R{\'{i}}o and Mark Wiebe and Pearu Peterson and Pierre
                 G{\'{e}}rard-Marchant and Kevin Sheppard and Tyler Reddy and
                 Warren Weckesser and Hameer Abbasi and Christoph Gohlke and
                 Travis E. Oliphant},
 year          = {2020},
 month         = sep,
 journal       = {Nature},
 volume        = {585},
 number        = {7825},
 pages         = {357--362},
 doi           = {10.1038/s41586-020-2649-2},
 publisher     = {Springer Science and Business Media {LLC}},
 url           = {https://doi.org/10.1038/s41586-020-2649-2}
}

@InProceedings{ pandas1,
  author    = { {W}es {M}c{K}inney },
  title     = { {D}ata {S}tructures for {S}tatistical {C}omputing in {P}ython },
  booktitle = { {P}roceedings of the 9th {P}ython in {S}cience {C}onference },
  pages     = { 56 - 61 },
  year      = { 2010 },
  editor    = { {S}t\'efan van der {W}alt and {J}arrod {M}illman },
  doi       = { 10.25080/Majora-92bf1922-00a }
}

@misc{pandas2,
  author       = {The pandas development team},
  title        = {pandas-dev/pandas: Pandas},
  year         = {2025},
  month        = jul,
  howpublished = {Zenodo},
  version      = {v2.3.1},
  doi          = {10.5281/zenodo.15831829},
  url          = {https://doi.org/10.5281/zenodo.15831829},
  note         = {Software release. SWHID: swh:1:dir:cf105b0362799b452aef6e27eaa64e292adcdfbb; 
                  origin=https://doi.org/10.5281/zenodo.3509134; 
                  visit=swh:1:snp:785dfb507f5191cf78b7ce5f86d728a8d579b2a0; 
                  anchor=swh:1:rel:9676a0e9ee3a3b294e775931e6d25d2a948f1697; 
                  path=pandas-dev-pandas-6d1e0b7},
}

@article{astropy:2013,
Adsurl = {http://adsabs.harvard.edu/abs/2013A%26A...558A..33A},
Archiveprefix = {arXiv},
Author = {{Astropy Collaboration} and {Robitaille}, T.~P. and {Tollerud}, E.~J. and {Greenfield}, P. and {Droettboom}, M. and {Bray}, E. and {Aldcroft}, T. and {Davis}, M. and {Ginsburg}, A. and {Price-Whelan}, A.~M. and {Kerzendorf}, W.~E. and {Conley}, A. and {Crighton}, N. and {Barbary}, K. and {Muna}, D. and {Ferguson}, H. and {Grollier}, F. and {Parikh}, M.~M. and {Nair}, P.~H. and {Unther}, H.~M. and {Deil}, C. and {Woillez}, J. and {Conseil}, S. and {Kramer}, R. and {Turner}, J.~E.~H. and {Singer}, L. and {Fox}, R. and {Weaver}, B.~A. and {Zabalza}, V. and {Edwards}, Z.~I. and {Azalee Bostroem}, K. and {Burke}, D.~J. and {Casey}, A.~R. and {Crawford}, S.~M. and {Dencheva}, N. and {Ely}, J. and {Jenness}, T. and {Labrie}, K. and {Lim}, P.~L. and {Pierfederici}, F. and {Pontzen}, A. and {Ptak}, A. and {Refsdal}, B. and {Servillat}, M. and {Streicher}, O.},
Doi = {10.1051/0004-6361/201322068},
Eid = {A33},
Eprint = {1307.6212},
Journal = {\aap},
Month = oct,
Pages = {A33},
Primaryclass = {astro-ph.IM},
Title = {{Astropy: A community Python package for astronomy}},
Volume = 558,
Year = 2013}

@ARTICLE{astropy:2018,
       author = {{Astropy Collaboration} and {Price-Whelan}, A.~M. and
         {Sip{\H{o}}cz}, B.~M. and {G{\"u}nther}, H.~M. and {Lim}, P.~L. and
         {Crawford}, S.~M. and {Conseil}, S. and {Shupe}, D.~L. and
         {Craig}, M.~W. and {Dencheva}, N. and {Ginsburg}, A. and {Vand
        erPlas}, J.~T. and {Bradley}, L.~D. and {P{\'e}rez-Su{\'a}rez}, D. and
         {de Val-Borro}, M. and {Aldcroft}, T.~L. and {Cruz}, K.~L. and
         {Robitaille}, T.~P. and {Tollerud}, E.~J. and {Ardelean}, C. and
         {Babej}, T. and {Bach}, Y.~P. and {Bachetti}, M. and {Bakanov}, A.~V. and
         {Bamford}, S.~P. and {Barentsen}, G. and {Barmby}, P. and
         {Baumbach}, A. and {Berry}, K.~L. and {Biscani}, F. and {Boquien}, M. and
         {Bostroem}, K.~A. and {Bouma}, L.~G. and {Brammer}, G.~B. and
         {Bray}, E.~M. and {Breytenbach}, H. and {Buddelmeijer}, H. and
         {Burke}, D.~J. and {Calderone}, G. and {Cano Rodr{\'\i}guez}, J.~L. and
         {Cara}, M. and {Cardoso}, J.~V.~M. and {Cheedella}, S. and {Copin}, Y. and
         {Corrales}, L. and {Crichton}, D. and {D'Avella}, D. and {Deil}, C. and
         {Depagne}, {\'E}. and {Dietrich}, J.~P. and {Donath}, A. and
         {Droettboom}, M. and {Earl}, N. and {Erben}, T. and {Fabbro}, S. and
         {Ferreira}, L.~A. and {Finethy}, T. and {Fox}, R.~T. and
         {Garrison}, L.~H. and {Gibbons}, S.~L.~J. and {Goldstein}, D.~A. and
         {Gommers}, R. and {Greco}, J.~P. and {Greenfield}, P. and
         {Groener}, A.~M. and {Grollier}, F. and {Hagen}, A. and {Hirst}, P. and
         {Homeier}, D. and {Horton}, A.~J. and {Hosseinzadeh}, G. and {Hu}, L. and
         {Hunkeler}, J.~S. and {Ivezi{\'c}}, {\v{Z}}. and {Jain}, A. and
         {Jenness}, T. and {Kanarek}, G. and {Kendrew}, S. and {Kern}, N.~S. and
         {Kerzendorf}, W.~E. and {Khvalko}, A. and {King}, J. and {Kirkby}, D. and
         {Kulkarni}, A.~M. and {Kumar}, A. and {Lee}, A. and {Lenz}, D. and
         {Littlefair}, S.~P. and {Ma}, Z. and {Macleod}, D.~M. and
         {Mastropietro}, M. and {McCully}, C. and {Montagnac}, S. and
         {Morris}, B.~M. and {Mueller}, M. and {Mumford}, S.~J. and {Muna}, D. and
         {Murphy}, N.~A. and {Nelson}, S. and {Nguyen}, G.~H. and
         {Ninan}, J.~P. and {N{\"o}the}, M. and {Ogaz}, S. and {Oh}, S. and
         {Parejko}, J.~K. and {Parley}, N. and {Pascual}, S. and {Patil}, R. and
         {Patil}, A.~A. and {Plunkett}, A.~L. and {Prochaska}, J.~X. and
         {Rastogi}, T. and {Reddy Janga}, V. and {Sabater}, J. and
         {Sakurikar}, P. and {Seifert}, M. and {Sherbert}, L.~E. and
         {Sherwood-Taylor}, H. and {Shih}, A.~Y. and {Sick}, J. and
         {Silbiger}, M.~T. and {Singanamalla}, S. and {Singer}, L.~P. and
         {Sladen}, P.~H. and {Sooley}, K.~A. and {Sornarajah}, S. and
         {Streicher}, O. and {Teuben}, P. and {Thomas}, S.~W. and
         {Tremblay}, G.~R. and {Turner}, J.~E.~H. and {Terr{\'o}n}, V. and
         {van Kerkwijk}, M.~H. and {de la Vega}, A. and {Watkins}, L.~L. and
         {Weaver}, B.~A. and {Whitmore}, J.~B. and {Woillez}, J. and
         {Zabalza}, V. and {Astropy Contributors}},
        title = "{The Astropy Project: Building an Open-science Project and Status of the v2.0 Core Package}",
      journal = {\aj},
         year = 2018,
        month = sep,
       volume = {156},
       number = {3},
          eid = {123},
        pages = {123},
          doi = {10.3847/1538-3881/aabc4f},
archivePrefix = {arXiv},
       eprint = {1801.02634},
 primaryClass = {astro-ph.IM},
       adsurl = {https://ui.adsabs.harvard.edu/abs/2018AJ....156..123A}
}

@ARTICLE{astropy:2022,
       author = {{Astropy Collaboration} and {Price-Whelan}, Adrian M. and {Lim}, Pey Lian and {Earl}, Nicholas and {Starkman}, Nathaniel and {Bradley}, Larry and {Shupe}, David L. and {Patil}, Aarya A. and {Corrales}, Lia and {Brasseur}, C.~E. and {N{"o}the}, Maximilian and {Donath}, Axel and {Tollerud}, Erik and {Morris}, Brett M. and {Ginsburg}, Adam and {Vaher}, Eero and {Weaver}, Benjamin A. and {Tocknell}, James and {Jamieson}, William and {van Kerkwijk}, Marten H. and {Robitaille}, Thomas P. and {Merry}, Bruce and {Bachetti}, Matteo and {G{"u}nther}, H. Moritz and {Aldcroft}, Thomas L. and {Alvarado-Montes}, Jaime A. and {Archibald}, Anne M. and {B{'o}di}, Attila and {Bapat}, Shreyas and {Barentsen}, Geert and {Baz{'a}n}, Juanjo and {Biswas}, Manish and {Boquien}, M{'e}d{'e}ric and {Burke}, D.~J. and {Cara}, Daria and {Cara}, Mihai and {Conroy}, Kyle E. and {Conseil}, Simon and {Craig}, Matthew W. and {Cross}, Robert M. and {Cruz}, Kelle L. and {D'Eugenio}, Francesco and {Dencheva}, Nadia and {Devillepoix}, Hadrien A.~R. and {Dietrich}, J{"o}rg P. and {Eigenbrot}, Arthur Davis and {Erben}, Thomas and {Ferreira}, Leonardo and {Foreman-Mackey}, Daniel and {Fox}, Ryan and {Freij}, Nabil and {Garg}, Suyog and {Geda}, Robel and {Glattly}, Lauren and {Gondhalekar}, Yash and {Gordon}, Karl D. and {Grant}, David and {Greenfield}, Perry and {Groener}, Austen M. and {Guest}, Steve and {Gurovich}, Sebastian and {Handberg}, Rasmus and {Hart}, Akeem and {Hatfield-Dodds}, Zac and {Homeier}, Derek and {Hosseinzadeh}, Griffin and {Jenness}, Tim and {Jones}, Craig K. and {Joseph}, Prajwel and {Kalmbach}, J. Bryce and {Karamehmetoglu}, Emir and {Ka{l}uszy{'n}ski}, Miko{l}aj and {Kelley}, Michael S.~P. and {Kern}, Nicholas and {Kerzendorf}, Wolfgang E. and {Koch}, Eric W. and {Kulumani}, Shankar and {Lee}, Antony and {Ly}, Chun and {Ma}, Zhiyuan and {MacBride}, Conor and {Maljaars}, Jakob M. and {Muna}, Demitri and {Murphy}, N.~A. and {Norman}, Henrik and {O'Steen}, Richard and {Oman}, Kyle A. and {Pacifici}, Camilla and {Pascual}, Sergio and {Pascual-Granado}, J. and {Patil}, Rohit R. and {Perren}, Gabriel I. and {Pickering}, Timothy E. and {Rastogi}, Tanuj and {Roulston}, Benjamin R. and {Ryan}, Daniel F. and {Rykoff}, Eli S. and {Sabater}, Jose and {Sakurikar}, Parikshit and {Salgado}, Jes{'u}s and {Sanghi}, Aniket and {Saunders}, Nicholas and {Savchenko}, Volodymyr and {Schwardt}, Ludwig and {Seifert-Eckert}, Michael and {Shih}, Albert Y. and {Jain}, Anany Shrey and {Shukla}, Gyanendra and {Sick}, Jonathan and {Simpson}, Chris and {Singanamalla}, Sudheesh and {Singer}, Leo P. and {Singhal}, Jaladh and {Sinha}, Manodeep and {Sip{H{o}}cz}, Brigitta M. and {Spitler}, Lee R. and {Stansby}, David and {Streicher}, Ole and {{{S}}umak}, Jani and {Swinbank}, John D. and {Taranu}, Dan S. and {Tewary}, Nikita and {Tremblay}, Grant R. and {Val-Borro}, Miguel de and {Van Kooten}, Samuel J. and {Vasovi{'c}}, Zlatan and {Verma}, Shresth and {de Miranda Cardoso}, Jos{'e} Vin{'i}cius and {Williams}, Peter K.~G. and {Wilson}, Tom J. and {Winkel}, Benjamin and {Wood-Vasey}, W.~M. and {Xue}, Rui and {Yoachim}, Peter and {Zhang}, Chen and {Zonca}, Andrea and {Astropy Project Contributors}},
        title = "{The Astropy Project: Sustaining and Growing a Community-oriented Open-source Project and the Latest Major Release (v5.0) of the Core Package}",
      journal = {\apj},
         year = 2022,
        month = aug,
       volume = {935},
       number = {2},
          eid = {167},
        pages = {167},
          doi = {10.3847/1538-4357/ac7c74},
archivePrefix = {arXiv},
       eprint = {2206.14220},
 primaryClass = {astro-ph.IM},
       adsurl = {https://ui.adsabs.harvard.edu/abs/2022ApJ...935..167A}
}

@article{scikit-learn,
  title={Scikit-learn: Machine Learning in {P}ython},
  author={Pedregosa, F. and Varoquaux, G. and Gramfort, A. and Michel, V.
          and Thirion, B. and Grisel, O. and Blondel, M. and Prettenhofer, P.
          and Weiss, R. and Dubourg, V. and Vanderplas, J. and Passos, A. and
          Cournapeau, D. and Brucher, M. and Perrot, M. and Duchesnay, E.},
  journal={Journal of Machine Learning Research},
  volume={12},
  pages={2825--2830},
  year={2011}
}

@misc{sergey_koposov_2024_12537467,
  author       = {Koposov, Sergey and
                  Speagle, Josh and
                  Barbary, Kyle and
                  Ashton, Gregory and
                  Bennett, Ed and
                  Buchner, Johannes and
                  Scheffler, Carl and
                  Cook, Ben and
                  Talbot, Colm and
                  Guillochon, James and
                  Cubillos, Patricio and
                  Asensio Ramos, Andr{\'e}s and
                  Dartiailh, Matthieu and
                  Ilya and
                  Tollerud, Erik and
                  Lang, Dustin and
                  Johnson, Ben and
                  Mendel, J. T. and
                  Higson, Edward and
                  Vandal, Thomas and
                  Daylan, Tansu and
                  Angus, Ruth and
                  Patel, R. and
                  Cargile, Phillip and
                  Sheehan, Patrick and
                  Pitkin, Matt and
                  Kirk, Matthew and
                  Leja, Joel and
                  Zuntz, Joe and
                  Goldstein, Danny},
  title        = {dynesty: v2.1.4},
  howpublished = {Zenodo},
  month        = jun,
  year         = 2024,
  doi          = {10.5281/zenodo.12537467},
  url          = {https://doi.org/10.5281/zenodo.12537467},
  note         = {Version v2.1.4},
}

@ARTICLE{2020SciPy-NMeth,
  author  = {Virtanen, Pauli and Gommers, Ralf and Oliphant, Travis E. and
            Haberland, Matt and Reddy, Tyler and Cournapeau, David and
            Burovski, Evgeni and Peterson, Pearu and Weckesser, Warren and
            Bright, Jonathan and {van der Walt}, St{\'e}fan J. and
            Brett, Matthew and Wilson, Joshua and Millman, K. Jarrod and
            Mayorov, Nikolay and Nelson, Andrew R. J. and Jones, Eric and
            Kern, Robert and Larson, Eric and Carey, C J and
            Polat, {\.I}lhan and Feng, Yu and Moore, Eric W. and
            {VanderPlas}, Jake and Laxalde, Denis and Perktold, Josef and
            Cimrman, Robert and Henriksen, Ian and Quintero, E. A. and
            Harris, Charles R. and Archibald, Anne M. and
            Ribeiro, Ant{\^o}nio H. and Pedregosa, Fabian and
            {van Mulbregt}, Paul and {SciPy 1.0 Contributors}},
  title   = {{{SciPy} 1.0: Fundamental Algorithms for Scientific
            Computing in Python}},
  journal = {Nature Methods},
  year    = {2020},
  volume  = {17},
  pages   = {261--272},
  adsurl  = {https://rdcu.be/b08Wh},
  doi     = {10.1038/s41592-019-0686-2},
}

@Article{Hunter2007,
  Author    = {Hunter, J. D.},
  Title     = {Matplotlib: A 2D graphics environment},
  Journal   = {Computing in Science \& Engineering},
  Volume    = {9},
  Number    = {3},
  Pages     = {90--95},
  publisher = {IEEE COMPUTER SOC},
  doi       = {10.1109/MCSE.2007.55},
  year      = 2007
}

@article{hasegawa2022,
  title = {Determining {{Dispersal Mechanisms}} of {{Protoplanetary Disks Using Accretion}} and {{Wind Mass Loss Rates}}},
  author = {Hasegawa, Yasuhiro and Haworth, Thomas J. and Hoadley, Keri and Kim, Jinyoung Serena and Goto, Hina and Juzikenaite, Aine and Turner, Neal J. and Pascucci, Ilaria and Hamden, Erika T.},
  year = {2022},
  month = feb,
  journal = {The Astrophysical Journal Letters},
  volume = {926},
  number = {2},
  pages = {L23},
  publisher = {The American Astronomical Society},
  issn = {2041-8205},
  doi = {10.3847/2041-8213/ac50aa},
  urldate = {2025-08-12},
  langid = {english}
}

@article{alexander2014,
  title = {The {{Dispersal}} of {{Protoplanetary Disks}}},
  author = {Alexander, R. and Pascucci, I. and Andrews, S. and Armitage, P. and Cieza, L.},
  year = {2014},
  journal = {Protostars and Planets VI, Henrik Beuther, Ralf S. Klessen, Cornelis P. Dullemond, and Thomas Henning (eds.), University of Arizona Press, Tucson, p.475-496},
  pages = {475},
  doi = {10.2458/azu_uapress_9780816531240-ch021},
  urldate = {2025-08-12},
  langid = {english}
}

@article{nesvorny2023,
  title = {Radial Distribution of Distant Trans-{{Neptunian}} Objects Points to {{Sun}}'s Formation in a Stellar Cluster},
  author = {Nesvorn{\'y}, David and Bernardinelli, Pedro and Vokrouhlick{\'y}, David and Batygin, Konstantin},
  year = {2023},
  month = dec,
  journal = {Icarus},
  volume = {406},
  pages = {115738},
  publisher = {Elsevier},
  issn = {0019-1035},
  doi = {10.1016/j.icarus.2023.115738},
  urldate = {2025-08-12}
}

@article{wyatt2017,
  title = {How to Design a Planetary System for Different Scattering Outcomes: Giant Impact Sweet Spot, Maximizing Exocomets, Scattered Discs},
  shorttitle = {How to Design a Planetary System for Different Scattering Outcomes},
  author = {Wyatt, M. C. and Bonsor, A. and Jackson, A. P. and Marino, S. and Shannon, A.},
  year = {2017},
  month = jan,
  journal = {Monthly Notices of the Royal Astronomical Society},
  volume = {464},
  number = {3},
  pages = {3385--3407},
  issn = {0035-8711},
  doi = {10.1093/mnras/stw2633},
  urldate = {2025-01-24}
}

@misc{puzia2025,
  title = {Spectral {{Characteristics}} of {{Interstellar Object 3I}}/{{ATLAS}} from {{SOAR Observations}}},
  author = {Puzia, Thomas H. and Rahatgaonkar, Rohan and Carvajal, Juan Pablo and Nayak, Prasanta K. and Luco, Baltasar},
  year = {2025},
  month = aug,
  number = {arXiv:2508.02777},
  eprint = {2508.02777},
  primaryclass = {astro-ph},
  publisher = {arXiv},
  doi = {10.48550/arXiv.2508.02777},
  urldate = {2025-08-24},
  archiveprefix = {arXiv}
}

@misc{yang2025,
  title = {Spectroscopic {{Characterization}} of {{Interstellar Object 3I}}/{{ATLAS}}: {{Water Ice}} in the {{Coma}}},
  shorttitle = {Spectroscopic {{Characterization}} of {{Interstellar Object 3I}}/{{ATLAS}}},
  author = {Yang, Bin and Meech, Karen J. and Connelley, Michael and Keane, Jacqueline V.},
  year = {2025},
  month = jul,
  number = {arXiv:2507.14916},
  eprint = {2507.14916},
  primaryclass = {astro-ph},
  publisher = {arXiv},
  doi = {10.48550/arXiv.2507.14916},
  urldate = {2025-08-24},
  archiveprefix = {arXiv}
}

@INPROCEEDINGS{pp7,
       author = {{Dr{\k{a}}{\.z}kowska}, J. and {Bitsch}, B. and {Lambrechts}, M. and {Mulders}, G.~D. and {Harsono}, D. and {Vazan}, A. and {Liu}, B. and {Ormel}, C.~W. and {Kretke}, K. and {Morbidelli}, A.},
        title = "{Planet Formation Theory in the Era of ALMA and Kepler: from Pebbles to Exoplanets}",
    booktitle = {Protostars and Planets VII},
         year = 2023,
       editor = {{Inutsuka}, S. and {Aikawa}, Y. and {Muto}, T. and {Tomida}, K. and {Tamura}, M.},
       series = {Astronomical Society of the Pacific Conference Series},
       volume = {534},
        month = jul,
        pages = {717},
          doi = {10.48550/arXiv.2203.09759},
archivePrefix = {arXiv},
       eprint = {2203.09759},
 primaryClass = {astro-ph.EP},
       adsurl = {https://ui.adsabs.harvard.edu/abs/2023ASPC..534..717D}
}

@article{pfalzner2024,
  title = {Trajectory of the Stellar Flyby That Shaped the Outer {{Solar System}}},
  author = {Pfalzner, Susanne and Govind, Amith and Portegies Zwart, Simon},
  year = {2024},
  month = nov,
  journal = {Nature Astronomy},
  volume = {8},
  pages = {1380--1386},
  issn = {2397-3366},
  doi = {10.1038/s41550-024-02349-x},
  urldate = {2025-09-30},
  langid = {english}
}

@article{fernandezFormationOortCloud1997,
  title = {The {{Formation}} of the {{Oort Cloud}} and the {{Primitive Galactic Environment}}},
  author = {Fern{\'a}ndez, Julio A.},
  year = {1997},
  month = sep,
  journal = {Icarus},
  volume = {129},
  number = {1},
  pages = {106--119},
  issn = {0019-1035},
  doi = {10.1006/icar.1997.5754},
  urldate = {2024-06-03}
}
\bibliographystyle{aasjournal}

\end{document}